# Security Audit of intel ICE Driver for e810 Network Interface Card

# LK049 – Bachelor of Science in Cyber Security and IT Forensics

# Project Report

Oisin O'Sullivan

24/03/2025

# Abstract

The security of enterprise-grade networking hardware and software is critical to ensuring the integrity, availability, and confidentiality of data in modern cloud and data center environments. Network interface controllers (NICs) play a pivotal role in high-performance computing and virtualization, although their privileged open access to system resources makes them a prime target for security vulnerabilities. This study presents a security analysis of the Intel ICE driver using the E810 Ethernet Controller, employing static analysis, fuzz testing, and timing-based side-channel evaluation to assess its robustness against exploitation.

The primary objective is to evaluate the driver's resilience against malformed inputs, identify architectural and implementation weaknesses, and determine whether timing discrepancies can be exploited for unauthorized inference of system states. The audit begins with a static code analysis, findings indicate that the lack of strict bounds checking, and the use of unsafe string operations could introduce security weaknesses. This is followed by fuzz testing, targeting driver components such as the Admin Queue, debugfs interface, and virtual function (VF) management subsystems. A combination of input mutation, command injection, and interface-aware fuzzing is applied to evaluate the driver's handling of anomalous inputs. The results demonstrate that the ICE driver exhibits strong input validation, preventing memory corruption, privilege escalation, and system crashes under standard fuzzing conditions.

Despite these protections, leveraging principals from *KernelSnitch,* the driver is susceptible to timing-based side-channel attacks [31]. By measuring execution time discrepancies in hash table lookups, an unprivileged attacker can infer VF occupancy states, potentially enabling network mapping attacks in multi-tenant cloud environments. Further investigation into the VF management subsystem highlights inefficiencies related to the Read-Copy-Update (RCU) synchronization mechanism, where the absence of explicit synchronization results in stale data persistence. This flaw increases the risk of memory leaks, stale pointer dereferencing, and out-of-memory (OOM) conditions under conditions of high VF churn. Kernel instrumentation confirms that occupied VF lookups complete significantly faster than unoccupied queries, reinforcing the potential for adversarial exploitation of timing channels.

| | |
|---|---|
| Name | **Oisin O'Sullivan** |
| Signature | *OO'S* |
| Date | 24/03/2025 |

iii

# Table of Contents

# List of Figures

# List of Tables

# Chapter 1: Introduction

Intel's Columbiaville E810 Ethernet controller is a product that was designed to meet the increasing need for high-speed data transmission and low-latency communication in enterprise and cloud environments [1]. This Ethernet controller powers large-scale infrastructures and ensures seamless data handling across servers, making it a critical component in the networking ecosystem. Figure 1.1 shows the block diagram for the Intel e810.

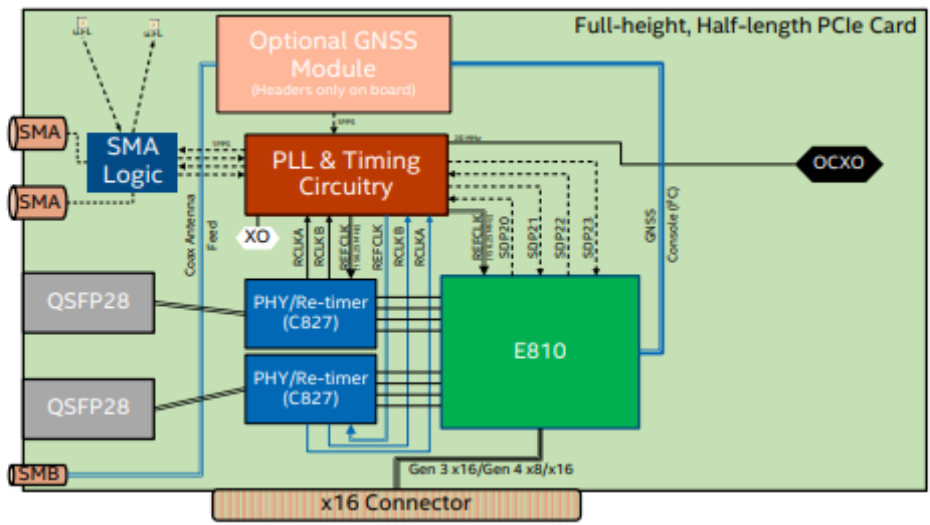

*Figure 1.1. E810-CQDA2T Block Diagram*

As with any system, security risks and potential vulnerabilities are inherent, especially within the device drivers that manage hardware-software interactions. Drivers in high-performance network controllers are often targets of security research due to their low-level access to system resources and critical role in data flow [2]. Vulnerabilities in these drivers could be exploited, leading to potential system compromises, unauthorised data access, or network disruptions. Ensuring the security of such drivers is essential for maintaining the integrity of modern networking infrastructure [2].

This project focuses on fuzzing the drivers for the Intel Columbiaville E810, specifically those available in the Intel Ethernet Linux ICE repository [3], [4]. Fuzzing is a dynamic testing technique that feeds random or semi-random inputs to software to trigger abnormal behaviour, such as crashes or memory corruption [5]. When applied to drivers, fuzzing can expose hidden security vulnerabilities that might otherwise go unnoticed during traditional testing [6], [7].

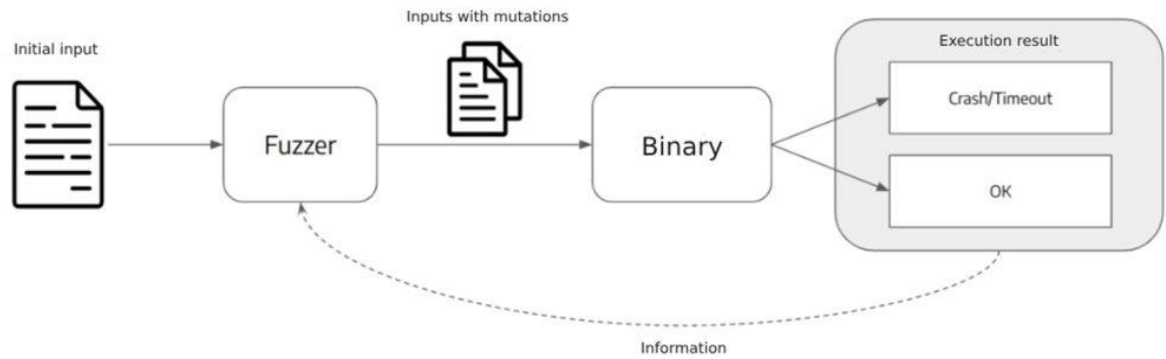

*Figure 1.2 -Simplified diagram of a fuzzer*

In addition to fuzzing, the project will explore the security features of the Intel Columbiaville E810 [3]. This will include an examination of data structures within the driver, which play a role in storing virtual functions, as part of SR-IOV [3], [13].

The primary objectives of this project are to identify potential weaknesses in the E810 drivers through fuzzing and to assess the efficacy of its built-in security features. By doing so, the project aims to contribute to the broader effort of securing enterprise networking hardware, which is increasingly becoming a target for sophisticated cyber threats. The research will also provide actionable insights for improving driver robustness and firmware protection in future hardware revisions.

## 1.1 Motivation for the project

The motivation for this project stems from my experience working at Intel, where I was involved in security architecture and research, working closely with architects and intel IPAS on matters related to vulnerability disclosure and mitigation. This background offered a unique perspective on the critical importance of securing hardware, especially enterprise-grade networking devices, in this case Intel's Columbiaville E810 Ethernet controller [1]. Due to legal constraints, the decision was made to test publicly available drivers [4]. A major benefit of choosing open-source drivers as the focus for this project is that they are publicly available [4] and concerns about inadvertent disclosure of proprietary or confidential IP are avoided.

Enterprise Security Challenges:

The necessity to mitigate potential vulnerabilities in network controllers is huge, due to the possibility of severe implications in the event of a viable compromise [12]. One of the critical insights is that even minor bugs in the device drivers could allow attackers to bypass system protections, especially since kernel drivers run with high levels of access and control over the system's resources [12]. These risks are particularly pertinent in data centres and cloud environments, where the Columbiaville E810 plays a key role in managing network traffic [1].

Fuzzing for Vulnerability Discovery:

This project will focus on fuzzing the drivers for the E810, as fuzzing is a proven method for uncovering hidden software bugs and vulnerabilities by generating unexpected input conditions [4], [11]. By using this technique, it simulates real-world attacks and identify any weaknesses in the drivers' code. The identification of these vulnerabilities is crucial for preventing potential exploitation that could lead to denial-of-service attacks, data leakage, or even remote code execution.

The importance of this research extends beyond just the security of the E810 controller itself. Networking controllers are foundational components in modern data centres, handling vast amounts of sensitive data traffic [12]. By securing these components, it increases the stability and security of critical infrastructures [11], [12]. This project, therefore, has the potential to contribute meaningfully to the broader efforts in improving security across the industry.

Most Recent Security Advisory from Intel

Intel's most recent security advisory highlights several vulnerabilities in the Driver of the Intel® Ethernet Adapters 800 Series Controllers, including the E810. As outlined in table 1, these vulnerabilities include denial-of-service attacks resulting from improper buffer restrictions, uncaught exceptions, and out-of-bounds reads. These issues emphasize the need for continuous security research and validation, as unpatched vulnerabilities can lead to significant impacts if exploited [16]. This makes the ICE Driver an ideal target for thorough testing and vulnerability discovery, justifying how this project will yield meaningful results.

| CVE ID | Description | CVSS Base Score | CVSS Vector | Impact |
|---|---|---|---|---|
| CVE-2021-0004 | Improper buffer restrictions in the firmware before version 1.5.3.0 may allow a privileged user to enable denial of service via local access. | 6.0 | CVSS:3.1/AV:L/AC:L/PR:H/UI:N/S:U/C:N/I:H/A:H | DoS |
| CVE-2021-0005 | Uncaught exception in the firmware before version 1.5.3.0 may allow a privileged user to enable denial of service via local access. | 6.0 | CVSS:3.1/AV:L/AC:L/PR:H/UI:N/S:U/C:N/I:H/A:H | DoS |
| CVE-2021-0006 | Improper conditions check in the firmware before version 1.5.4.0 may allow a privileged user to enable denial of service via local access. | 5.1 | CVSS:3.1/AV:L/AC:L/PR:H/UI:N/S:U/C:N/I:L/A:H | DoS |
| CVE-2021-0007 | Uncaught exception in the firmware before version 1.5.1.0 may allow a privileged attacker to enable denial of service via local access. | 4.4 | CVSS:3.1/AV:L/AC:L/PR:H/UI:N/S:U/C:N/I:N/A:H | DoS |
| CVE-2021-0008 | Uncontrolled resource consumption in the firmware before version 1.5.3.0 may allow a privileged user to enable denial of service via local access. | 4.4 | CVSS:3.1/AV:L/AC:L/PR:H/UI:N/S:U/C:N/I:N/A:H | DoS |
| CVE-2021-0009 | Out-of-bounds read in the firmware before version 1.5.3.0 may allow an unauthenticated user to enable denial of service via adjacent access. | 4.3 | CVSS:3.1/AV:A/AC:L/PR:N/UI:N/S:U/C:N/I:N/A:L | DoS |

*Table 1 - ""Intel® Ethernet E810 Adapter Driver Advisory""*

This table defines how the identified vulnerabilities in the Intel Ethernet Adapter make it a rich target for security testing, supporting the project's goal of uncovering and mitigating potential risks.

# Chapter 2: Background Theory

## 2.1. Fuzzing Techniques for Network Controllers

Fuzzing is particularly effective for software with complex interactions, such as device drivers, the application of fuzzing with Input-to-State Correspondence, showcases its potential to reveal subtle and critical vulnerabilities [5], [6], [7]. The effectiveness of fuzzing in exposing vulnerabilities in high-privilege software components highlights its importance for security researchers and developers [5], [6].

The fuzzing process involves generating diverse inputs and systematically testing the target software to observe how it handles unexpected data [5]. This method is valuable for assessing the robustness of all types of software, which manage critical data flows and interact closely with system hardware. By simulating a variety of abnormal conditions, fuzzing can expose hidden security flaws that may be exploited by attackers to compromise system integrity [5]. [6]. [7].

The rising complexity of enterprise-grade hardware, in this case, components like Intel's Columbiaville E810 Ethernet controllers, highlights the importance of securing these systems [3]. Vulnerabilities in their drivers can pose substantial risks, potentially leading to data breaches and service disruptions in enterprise environments.

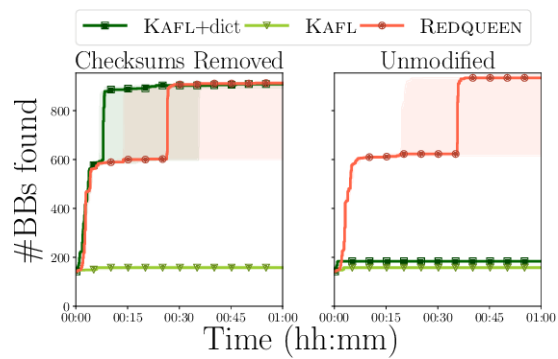

Figure 3.1 - Evaluating the impact of REDQUEEN vs KAFL.

## 2.2. Intel Ethernet 800 Series Controllers

The Intel Ethernet 800 Series, including the Columbiaville E810, represents a significant advancement in network controller technology, released in 2020 [4]. The open-source drivers available for these controllers allow for extensive scrutiny and modification, providing an opportunity for researchers to assess and improve their security [4].

The drivers for the Intel E810 are hosted on GitHub [4]. These drivers manage the interaction between the operating system and the hardware, improving the understanding of the architecture and functionality of these drivers is crucial for identifying potential weaknesses and enhancing their security posture [3].

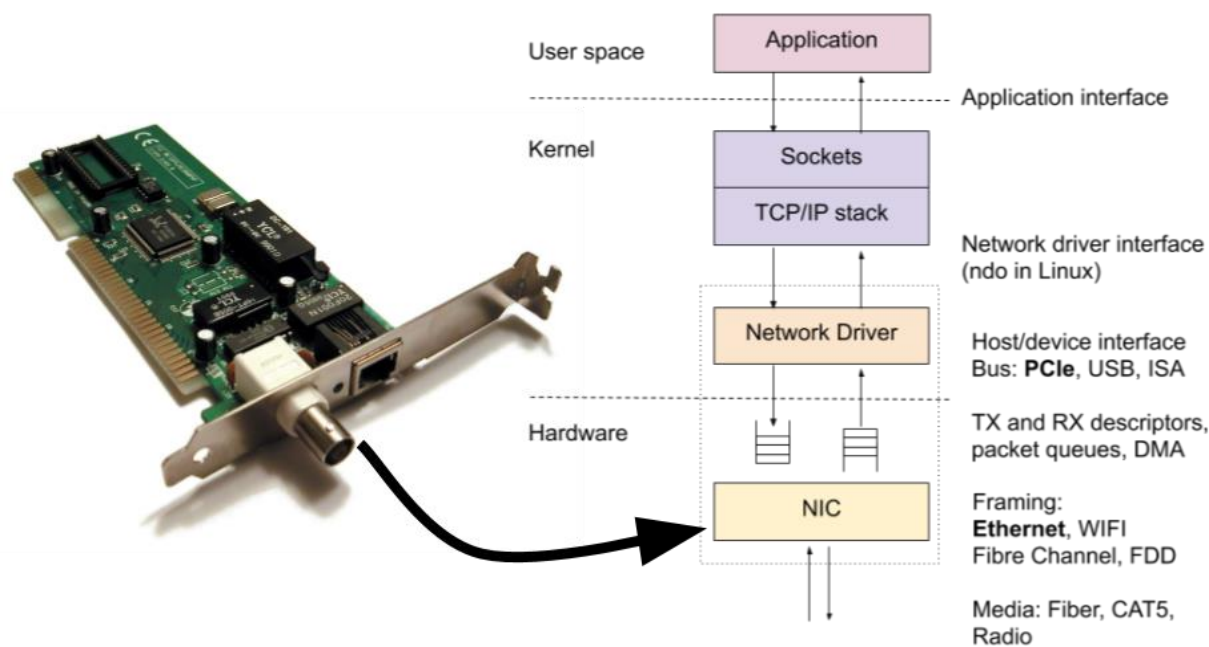

*Figure 2.2 - Example" Technical Deep Dive on NICs @IETF105"*

## 2.3. Driver Architecture: Physical and Virtual Function Drivers

Drivers play a critical role in managing the functionality of hardware components [10], and the distinction between Physical Function Drivers and Virtual Function Drivers is key to understanding potential risks in enterprise-grade networking hardware [10, [13].

**Physical Function Drivers** manage the hardware's most fundamental operations, but if improperly configured or exploited, they could severely disrupt the system [3], [13]. Privilege escalation or kernel-level vulnerabilities in these drivers can lead to extensive damage, potentially damaging the chip or corrupting other essential system functions [6], [10], [12]. The power that physical function drivers wield over the hardware means they are particularly dangerous if a vulnerability allows them to interfere with other components [3], [10], [13].

**Virtual Function Drivers** are intended to provide isolation in virtualized environments [3]. [13]. Their role is to ensure that each virtual machine (VM) on a host operates independently, preventing one VM from affecting another [13]. However, if a virtual function can break isolation and impact another VM on the same host, this could become an attractive target for this project, thus finding a vulnerability that compromises this isolation would be particularly valuable from a security research perspective [12], [13].

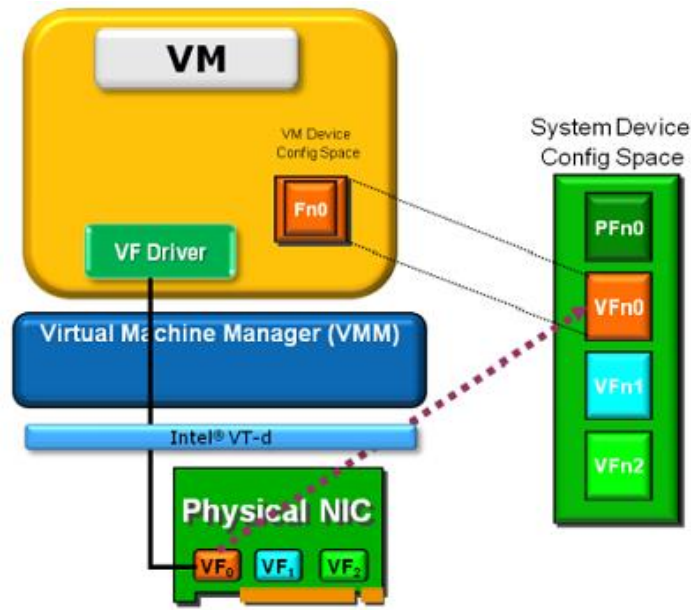

*Fig. 2.3 Configure SR-IOV and Create a Set of Virtual Functions*

## 2.4 Fuzzing Mechanisms

Fuzzing has become a critical method for uncovering software vulnerabilities by generating random inputs and inducing unexpected behaviours in systems. One prominent fuzzing approach, DIFUZE, is an interface-aware fuzzing tool designed to target kernel drivers. Corina et al [6]. demonstrated its ability to uncover previously unknown vulnerabilities in kernel drivers, highlighting the effectiveness of fuzzing techniques for enhancing hardware security [6]. The application of DIFUZE to the ice drivers of the Columbiaville E810 could provide valuable insights into their security weaknesses.

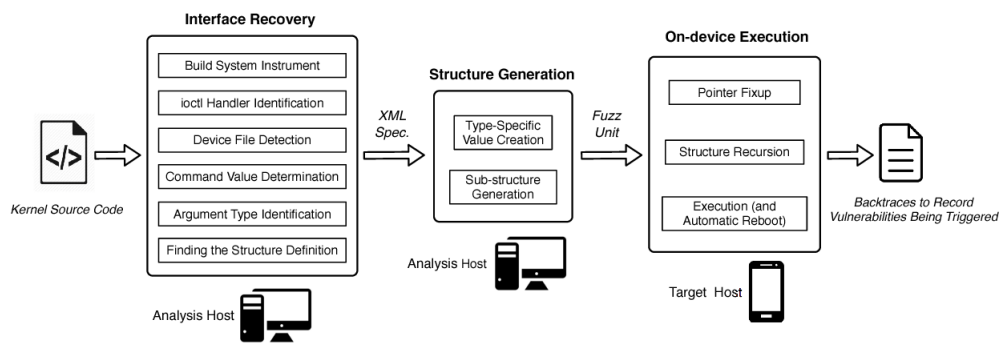

*Figure [2.4]" Figure 1: The DIFUZE approach diagram. DIFUZE analyses the provided kernel sources using a composition of analyses to extract driver interface information, such as valid ioctl commands and argument structure types. It synthesizes instances of these structures and dispatches them to the target device, which triggers ioctl execution with the given inputs and, eventually, finds crashes in the device drive*rs"

Another notable advancement in fuzzing is the REDQUEEN technique, which addresses common challenges in testing kernel drivers by focusing on input-to-state correspondence, demonstrated in Figure 3.1. This method has proven effective in identifying bugs in Operating System kernel drivers, making it a potential candidate for adaptation to the ice drivers of Intel's E810 Ethernet controllers [7]. Additionally, the kernel-AFL (kAFL) framework offers a hypervisor-based approach to coverage-guided fuzzing, enabling the discovery of vulnerabilities across various kernel components [8]. kAFL's success in detecting flaws within file system drivers suggests its applicability to improving the security of Intel's ice drivers [8].

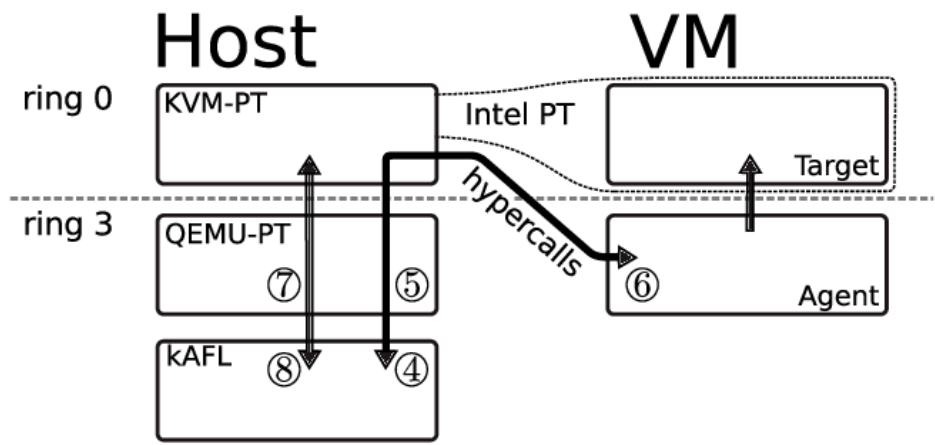

Figure 2.5  "High-level overview of the kAFL architecture"

## 2.5.       Challenges      in      Fuzzing      Device      Drivers

Despite the advancements in fuzzing methodologies since 2016, analysing device drivers remains a complex task [11]. Kernel drivers, particularly those responsible for hardware components, require sophisticated tools and frameworks to ensure thorough testing. USBFuzz, a tool designed for fuzzing USB device drivers, has been effective in identifying vulnerabilities across multiple operating systems [9]. Insights from USBFuzz could inform the development of new fuzzing techniques for the ice drivers of the E810, reinforcing the need for specialized approaches to uncover vulnerabilities in enterprise hardware [9], [10].

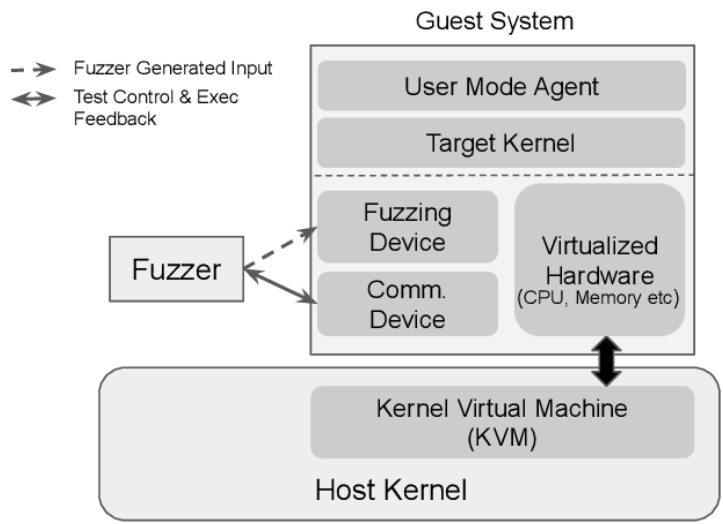

Figure 2.6 - "Overview of USBFuzz"

## 2.6. Addressing Knowledge Gaps

While existing research highlights the potential of various fuzzing techniques, there are significant knowledge gaps in their application to Intel's ice drivers. Few studies have specifically focused on adapting methods from REDQUEEN or kAFL to target the unique characteristics of the ice drivers in the Columbiaville E810 [3], [7], [8], [10].

Future research should address these gaps by tailoring fuzzing techniques to the intricacies of Intel's hardware and associated descriptor formats. While fuzzing has proven effective for discovering vulnerabilities, there is limited exploration of its integration with other security measures. Combining fuzzing with static analysis or dynamic testing would offer a more comprehensive security strategy, providing enhanced protection for enterprise-grade hardware like Intel's Columbiaville controllers [3]. Addressing the identified knowledge gaps through focused studies will be essential in strengthening the security of these components and enhancing the overall resilience of enterprise environments against emerging threats.

# Chapter 3: Flawfinder Process and Findings

## 3.1. Static Analysis

To begin, a security analysis of the ICE driver source code is conducted using Flawfinder, a static analysis tool designed to detect potential security vulnerabilities and weaknesses in C/C++ codebases [17]. Flawfinder is particularly effective for identifying issues such as buffer overflows and format string vulnerabilities [17], [20],[21]. Through this tool, a detailed review of the ICE driver's code was conducted, to detect patterns and coding practices that may lead to security risks.

After running Flawfinder on the ICE driver code, the resulting report had identified potential vulnerabilities. Among the findings were buffer overflow issues in the `ice_debug_cq` and `ice_parse_item_dflt` functions [20], within `ice_controlq.c` and `ice_parser.c` respectively [4]. This section outlines the vulnerability reports, including analyses of the affected code and recommendations for mitigation.

### 3.1.1 ice_debug_cq Function

A potential buffer overflow weakness was identified in the `ice_debug_cq` within `ice_controlq.c`, within the ICE driver code [4]. The weakness arises due to the use of `sprintf` for formatting without bounds checking [33]. This lack of boundary management risks writing beyond the allocated buffer size, leading to a potential overwrite of adjacent memory [33]. Buffer overflows are a critical concern in software security and have been extensively documented in literature [18], [19], [20], [33].

### 3.1.2 Affected Function

```c
static void ice_debug_cq(struct ice_hw *hw, struct ice_ctl_q_info *cq,
			 void *desc, void *buf, u16 buf_len, bool response)
{
	struct ice_aq_desc *cq_desc = desc;
	u16 datalen, flags;

	if (!IS_ENABLED(CONFIG_DYNAMIC_DEBUG) &&
	    !((ICE_DBG_AQ_DESC | ICE_DBG_AQ_DESC_BUF) & hw->debug_mask))
		return;

	if (!desc)
		return;

	datalen = le16_to_cpu(cq_desc->datalen);
	flags = le16_to_cpu(cq_desc->flags);

	ice_debug(hw, ICE_DBG_AQ_DESC, "%s %s: opcode 0x%04X, flags 0x%04X, datalen 0x%04X, retval 0x%04X\n\tcookie (h,l) 0x%08X 0x%08X\n\tparam (0,1)  0x%08X 0x%08X\n\taddr (h,l)   0x%08X 0x%08X\n",
		  ice_ctl_q_str(cq->qtype), response ? "Response" : "Command",
		  le16_to_cpu(cq_desc->opcode), flags, datalen,
		  le16_to_cpu(cq_desc->retval),
		  le32_to_cpu(cq_desc->cookie_high),
		  le32_to_cpu(cq_desc->cookie_low),
		  le32_to_cpu(cq_desc->params.generic.param0),
		  le32_to_cpu(cq_desc->params.generic.param1),
		  le32_to_cpu(cq_desc->params.generic.addr_high),
		  le32_to_cpu(cq_desc->params.generic.addr_low));
	/* Dump buffer iff 1) one exists and 2) is either a response indicated
	 * by the DD and/or CMP flag set or a command with the RD flag set.
	 */
	if (buf && cq_desc->datalen &&
	    (flags & (ICE_AQ_FLAG_DD | ICE_AQ_FLAG_CMP | ICE_AQ_FLAG_RD))) {
		char prefix[] = KBUILD_MODNAME " 0x12341234 0x12341234 ";

		sprintf(prefix, KBUILD_MODNAME " 0x%08X 0x%08X ",
			le32_to_cpu(cq_desc->params.generic.addr_high),
			le32_to_cpu(cq_desc->params.generic.addr_low));
		ice_debug_array_w_prefix(hw, ICE_DBG_AQ_DESC_BUF, prefix,
					 buf,
					 min_t(u16, buf_len, datalen));
	}
}
```

Fig 3.1 - screenshot of `ice_debug_cq` within `src/ice_controlq.c`

3.1.3 Weakness Description

The weakness stems from the use of `sprintf` to format data into the buffer prefix without confirming that the buffer size is adequate to store the formatted string [4], [17]. The prefix buffer is initially sized based on a default string:

- `char prefix[] = KBUILD_MODNAME " 0x12341234 0x12341234 ";`

However, it is later overwritten by `sprintf`, which could lead to memory overflow if the formatted string length exceeds the allocated buffer size:

```
sprintf(prefix, KBUILD_MODNAME " 0x%08X 0x%08X ",
        le32_to_cpu(cq_desc->params.generic.addr_high),
        le32_to_cpu(cq_desc->params.generic.addr_low));
```

In this instance however, `KBUILD_MODENAME`, is set upon compilation, and can only be changed in the `Makefile` pre-compilation, which is possible as per the below image, but in this case, the added bits are accounted for, and this cannot be changed after compilation [4]. Although you can modify binaries to change the contents of `KBUILD_MODNAME`, but this modified string must adhere to the original length at compilation.

Modified `KBUILD_MODNAME` Proofs:

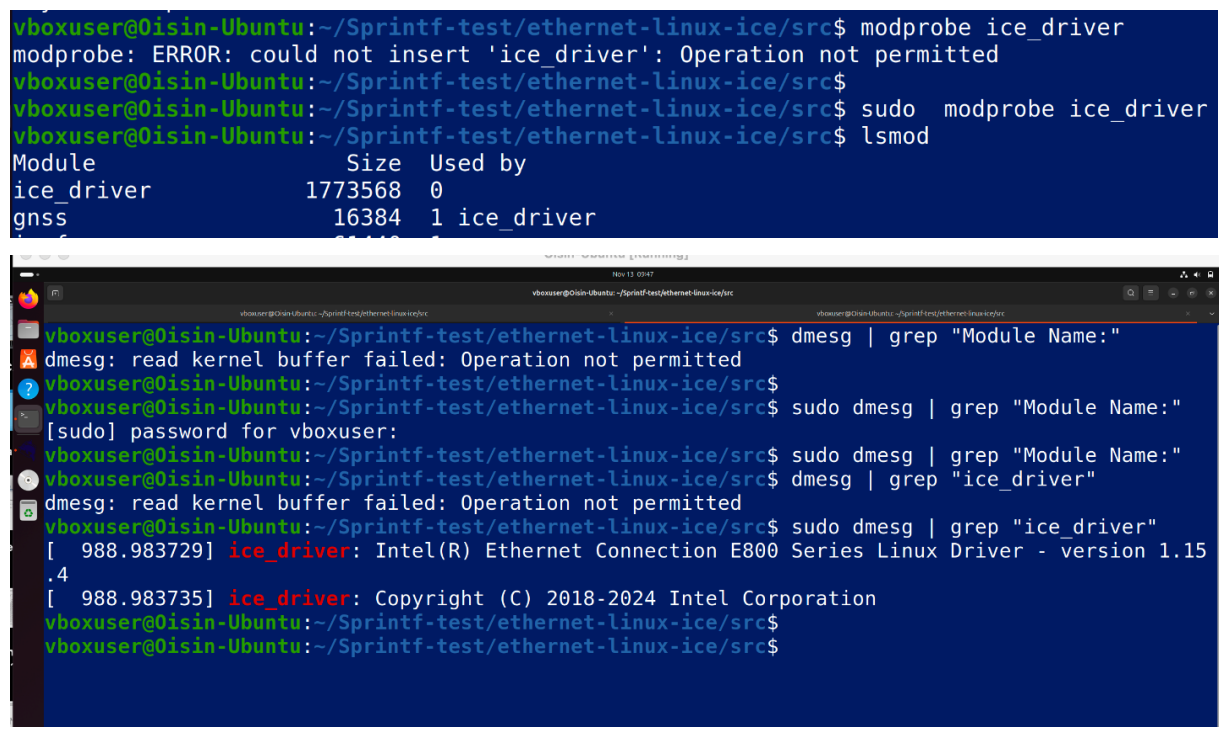

Fig 3.2 - screenshot of proof that `KBUILD_MODENAME` can be changed and not affect compilation (by just modifying `Makefile`).

### 3.1.4 Potential Impact

Broadly speaking, buffer overflows can lead to severe consequences, including arbitrary code execution, system instability, and privilege escalation [18], [20], [22]. If exploited, an attacker could use this overflow to manipulate the execution flow or corrupt sensitive data, compromising the system's integrity and security [33]. But in this case, the impact is little to none, as it involves modifying source code, and Linux has failsafe's to stop creating a name that's too large over 64 bytes [4].

## 3.2 ice_parse_item_dflt Function

A potential buffer overflow weakness was also discovered in the `ice_parse_item_dflt()` function located within the `ice_parser.c` source file of the ICE driver [4]. Like the previous issue, this weakness arises from inadequate bounds checking when copying data, which could lead to memory corruption and potential exploitation [33].

### 3.2.1 Affected Function

```
23  void ice_lbl_dump(struct ice_hw *hw, struct ice_lbl_item *item)
24  {
25      dev_info(ice_hw_to_dev(hw), "index = %d\n", item->idx);
26      dev_info(ice_hw_to_dev(hw), "label = %s\n", item->label);
27  }
28
29  void ice_parse_item_dflt(struct ice_hw *hw, u16 idx, void *item,
30                           void *data, int size)
31  {
32      memcpy(item, data, size);
33  }
```

Fig 3.3 - screenshot of `ice_parse_item_dflt()` function within the src/ice_parser.c

### 3.2.2 Weakness Description

The function `ice_parse_item_dflt()` uses `memcpy()` to copy data into item without validating that the destination buffer can safely hold the data being copied [4]. The absence of bounds checking allows for a buffer overflow if the size parameter exceeds the allocated size of item. This is a violation of secure coding practices regarding buffer management [19], [22].

### 3.2.3 Potential Impact

If exploited, this weakness could lead to memory corruption, system instability, or further exploitation opportunities that may compromise system security [21], [23]. An attacker could potentially manipulate the size parameter to overflow the buffer, leading to arbitrary code execution or denial of service.

## 3.3 Recommendations

To mitigate these weaknesses, I recommend the following actions:

### 3.3.1 Replace Unsafe Functions with Bounded Alternatives

By using `snprintf` [35], the function can limit the number of characters written, preventing overflow:

```
941            char prefix[] = KBUILD_MODNAME " 0x12341234 0x12341234 ";
942
943        snprintf(prefix, sizeof(prefix), KBUILD_MODNAME " 0x%08X 0x%08X ",
944            le32_to_cpu(cq_desc->params.generic.addr_high),
945            le32_to_cpu(cq_desc->params.generic.addr_low));
946        ice_debug_array_w_prefix(hw, ICE_DBG_AQ_DESC_BUF, prefix,
947                    buf,
948                    min_t(u16, buf_len, datalen));
949    }
950 }
```

Fig 3.4 - Screenshot of mitigation for `KBUILD_MODNAME` improper use of `sprintf` via `snprintf`.

This change ensures that the formatted string does not exceed the buffer's capacity [25].

```
void ice_parse_item_dflt(struct ice_hw *hw, u16 idx, void *item, void *data, int size, int item_size)
{
    if (size > item_size) {
        dev_err(ice_hw_to_dev(hw), "Error: Buffer overflow attempt detected. Size exceeds item buffer.\n");
        return;
    }
    memcpy(item, data, size);
}
```

Fig 3.5 - Screenshot of Mitigation located in `ice_parse_item_dflt()` function within the `src/ice_parser.c` via use of `memcpy_s`

Alternatively, use `memmove_s`, which includes built-in bounds checking [26].

# Chapter 4: Fuzzing Experiments

This chapter details the fuzzing experiments conducted on the Intel ICE driver for the E810 Ethernet Controller [4], focusing on its Admin Queue, debugfs interface, and runtime configuration mechanisms [30]. The objective is to assess the driver's robustness in defending against malformed input to prevent potential security weaknesses.

Three custom fuzzing scripts, `AdminQCmd-UI-Fuzz.py`, `Debugfs_inputV_Fuzz_test.py`, and `adqsetup-atheris-instrumentation.py`, were created to facilitate targeted fuzzing of the ICE driver. These scripts integrate various fuzzing methodologies inspired by "*Redqueen*", "*kAFL*", "*DIFUZE*", and USB fuzzing to supplement code coverage and explore edge-case behaviours in device driver interaction [6], [7], [8], [9].

The `AdminQCmd-UI-Fuzz.py` script tests the ICE driver's Admin Queue (AdminQ) interface, which handles privileged commands for managing NIC configurations [30]. This script employs a feedback-driven fuzzing approach inspired by Redqueen, where generated commands are analysed for anomalies, kernel log outputs, and system stability [7]. It also integrates kAFL-inspired multi-threading and logging mechanisms, enabling parallel execution of fuzzing tasks while capturing crashes, system errors, and unexpected behaviours [8].

The `Debugfs_inputV_Fuzz_test.py` script targets the debugfs interface of the ICE driver. Which is a common diagnostic and debugging mechanism in Linux [30], [41]. This script leverages a DIFUZE-like interface fuzzing to explore potential weaknesses in debugfs file handling [6]. It generates malicious edge-case payloads, including format string exploits, command injections, and buffer overflows, testing how the driver handles huge input written to debugfs files. Additionally, it integrates USB fuzzing techniques, historically used to test device driver robustness, by manipulating privileged file system interactions and monitoring unintended behaviour through system log analysis [9].

The `adqsetup-atheris-instrumentation.py` script is an instrumented version of the standard adqsetup utility used in the Intel ICE repository for Adaptive Queueing (ADQ) configuration [30]. This version has been modified to integrate Atheris, a fuzzing framework developed by Google for Python applications [32]. Atheris instrumentation allows string manipulation analysis, automated corpus mutation and input-to-state tracking. Which improves the efficiency of fuzzing queue configurations and command-line input handling [32]. By instrumenting functions within the script, Atheris enables deeper insights into how ADQ setup commands respond to malformed inputs, ensuring that validation mechanisms in the ICE driver's user-space tools are sufficient against unexpected or malicious input patterns.

All three scripts follow structured fuzzing methodologies:

- Command Injection & Input Mutation: Redqueen-inspired magic value mutation and automated corpus expansion are used to generate fuzzing inputs [7].

- Kernel Log & System Response Monitoring: Execution is continuously monitored using kAFL-like instrumentation, where kernel logs (`dmesg`, `journalctl`) are analyzed in real-time to detect anomalies [8].
- Crash & Anomaly Detection: Logs are scanned using regular expression-based pattern matching, focusing on kernel crashes, segmentation faults, driver failures, and security-related messages.
- File Integrity & Debugfs Monitoring: Debugfs fuzzing employs hash-based file integrity checks, detecting silent modifications that may indicate security vulnerabilities [41].
- Queue Configuration Fuzzing via Atheris: Atheris instrumentation allows for adaptive input mutation, improving the ability to detect subtle failures in ADQ command processing.

Due to project constraints and deadlines, kAFL, DIFUZE, or Redqueen were not employed in their entirety. However, various aspects of all were systematically integrated as their principles allowed me to create a custom fuzzing framework tailored to the specific characteristics of the Intel ICE driver using python. By leveraging feedback-driven input mutation from Redqueen, kernel-space instrumentation inspired by kAFL, and interface-aware fuzzing methodologies from DIFUZE, this approach effectively examines the security resilience of the driver. The experiments focus on the strength of untrusted input handling, command execution integrity, and system interaction security to identify potential vulnerabilities. The following sections provide a comprehensive analysis of each fuzzing experiment, detailing the methodology and execution results.

The experiment was conducted on a system running Ubuntu 22.04 LTS, equipped with Linux Kernel 6.8.0-generic. The system specifications are as follows:

- Processor: Intel Core i7-770 @ 3.6 GHz
- Memory: 16 GB DDR4 RAM
- Network Interface Card (NIC): Intel E810 Ethernet Adapter
- Operating System: Ubuntu 22.04 LTS
- Kernel Version: Linux 6.8.0-generic.
- Intel ICE Driver: v1.16.3

## 4.1 Admin Queue Command Fuzz Test

The AdminQcmd-Fuzz-Test.py script performs fuzz testing on the Admin Queue (AdminQ) interface of the ICE driver. The AdminQ is a critical component responsible for handling administrative commands related to device configuration, queue management, and firmware updates [4].

This test involves:

- Randomized input generation, where commands are constructed with arbitrary strings, special characters, and large inputs.

- Execution of fuzzed commands through adqsetup with various parameters.

- Real-time log monitoring, capturing errors such as kernel panics, segmentation faults, and driver failures.

- Detection of anomalies based on pre-defined patterns, such as "down", "segmentation faut" or "core dumped"

The objective of this test is to assess how the Admin Queue responds to unexpected inputs. If the driver fails to properly handle malformed requests, it may lead to denial-of-service conditions, unexpected reboots, or memory corruption [23].

Observed Impact on the ICE Driver

- Exposed potential weaknesses in input validation mechanisms.

- Identified instances where erroneous commands triggered system warnings.

- Highlighted cases where specific malformed inputs caused prolonged execution times, indicating possible performance degradation.

Results

Fig 4.1 - Adminq Fuzzing results screenshot

In this case, all inputs were handled gracefully as the system either accepted a valid command, or threw an error for an incorrect one, this demonstrates adequate input validation and protects against backtick execution, special character execution, and extra-long inputs.

## 4.2 Debugfs Input Validation Fuzz Test

The Debugfs_inputV_Fuzz_test.py script targets the debugfs interface of the ICE driver, which exposes internal driver state and logs. This interface provides valuable debugging information but can also become an attack surface if not properly secured.

This test:

- Writes malformed inputs to debugfs files to test for buffer overflows and improper handling.

- Monitors file integrity by generating hash signatures before and after modifications.

- Attempts various exploit techniques, including:

    o   Null byte injection to test for improper termination handling.

    o   Format string exploits that could expose memory contents.

    o   Command injection attempts to evaluate privilege separation.

Observed Impact on the ICE Driver

- Highlighted potential input validation gaps in the ICE debugfs interface.

- Exposed scenarios where writing malformed payloads resulted in unexpected driver behaviours.

- Suggested that certain debugfs paths might be more vulnerable to user manipulation than others.

Results

Fig 4.2 - Debugfs Fuzzing results screenshot

In this instance, the test found no unexpected behaviour, as "Command's executed" notate commands accepted by the systems and "error's" listen out for faults like "core dumped" & "DOWN".

### 4.3 Atheris Instrumentation-Based Fuzz Test

The adqsetup-atheris-instrumentation.py script takes a different approach by incorporating Atheris, a fuzzing framework that instruments Python-based configuration handling [36]. Instead of targeting binary execution paths, this test focuses on detecting vulnerabilities in the ADQ (Adaptive Queueing) configuration mechanism [30].

Key aspects of this test include:

- Regex and string input analysis, ensuring that unexpected values do not trigger unexpected behaviours.

- Instrumentation of internal functions to detect improper memory handling.

- Adaptive input mutation, where fuzzing payloads evolve based on prior outcomes to increase efficiency.

The primary goal is to determine whether malformed ADQ configurations could lead to system instability.

Observed Impact on the ICE Driver

- Indicated that certain malformed configurations could lead to unexpected queue behaviours.

- Suggested areas where additional validation could improve driver robustness.

- Showed that regex-based processing of configurations is a potential attack vector.

Results

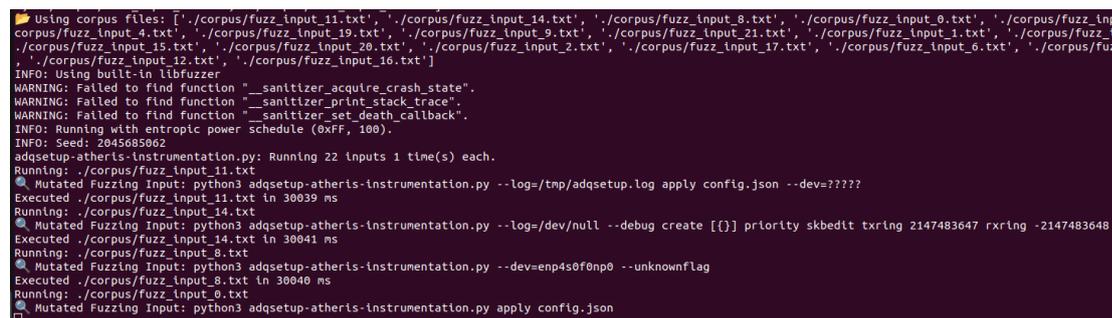

Fig 4.3 - Arteris Fuzzing results screenshot.

Here's an image of the Atheris test, where a corpus of valid inputs must be supplied, of which the fuzzer mutates these inputs for the purpose of causing unexpected behaviour, this yielded nothing unintended and all malformed inputs the fuzzer generated were handled correctly after ~6 Hours of Fuzzing.

## 4.4 Conclusions

This chapter presented an analysis of fuzzing experiments conducted on the Intel ICE driver for the E810 Ethernet Controller, focusing on its Admin Queue interface, debugfs interactions, and ADQ configuration mechanisms [30]. The primary objective was to assess the driver's resilience against malformed input and identify potential security weaknesses. Three custom fuzzing scripts: `AdminQCmd-UI-Fuzz.py`, `Debugfs_inputV_Fuzz_test.py`, and `adqsetup-atheris-instrumentation.py`. Which were developed to systematically test different components of the driver using methodologies inspired by Redqueen, kAFL, DIFUZE, and USB fuzzing [6], [7], [8], [9].

Despite integrating various fuzzing techniques, the experiments did not uncover critical security flaws in the ICE driver. The Admin Queue and debugfs interfaces demonstrated robust input validation, preventing crashes, privilege escalation, or memory corruption. Additionally, Atheris-based fuzzing of ADQ configurations confirmed that malformed inputs were appropriately handled. These findings suggest that the ICE driver exhibits strong resilience to fuzzing-based attacks, reinforcing the reliability of its input handling mechanisms from user-space.

# Chapter 5: Virtual Function Side-Channel Hash Table Tests

## 5.1 Introduction: Virtual Functions and Side-Channel Attacks

In multi-tenant cloud environments, secure management of network resources is essential to ensure fair allocation and security across different virtualized workloads [12]. Network Interface Cards (NICs) that support Single Root I/O Virtualization (SR-IOV) allow multiple Virtual Functions (VFs) to share the same physical hardware, each functioning as an independent virtualized network device [4],[13],[30]. The Intel ICE driver, which supports high-performance network operations, utilizes a hash table to manage these VF IDs efficiently. This hash table structure enables rapid lookup and access to VF information, optimizing resource allocation and packet processing [4], [12], [30], [43].

Leveraging principles from *"Kernelsnitch"*, a methodology designed for detecting kernel-level timing vulnerabilities, this study extends to investigate the feasibility of timing-based inference attacks against the ICE driver's hash table implementation [31], [43]. The objective is to determine whether an attacker can deduce VF presence using user-space tools and whether timing variations in hash table lookups expose underlying kernel state information [31], [43].

A hash table in the context of the ICE driver, is used to hold the VF IDs, thus the implementation of said hash table could introduce the potential for timing-based side-channel vulnerabilities [4], [43]. Since hash table lookups exhibit variable execution times based on occupancy, load factor, and hash collisions, an attacker with sufficient access to timing information could infer whether a given VF ID is active [31]. This information leakage can be exploited in various ways, such as targeted denial-of-service (DoS) attacks against active tenants or the intentional creation of hash collisions to degrade network performance [2].

The experiment was conducted on a system running Ubuntu 22.04 LTS, equipped with Linux Kernel 6.8.0-generic. The system specifications are as follows:

- Processor: Intel Core i7-770 @ 3.6 GHz
- Memory: 16 GB DDR4 RAM
- Network Interface Card (NIC): Intel E810 Ethernet Adapter
- Operating System: Ubuntu 24.04 LTS
- Kernel Version: Linux 6.11.0-generic.
- Intel ICE Driver: v1.16.3

## 5.2 User Space Fuzzing-Based Enumeration of VF IDs

To evaluate the potential for user-space inference attacks, I first examined whether common system utilities could reliably distinguish between occupied and unoccupied VF IDs. By leveraging fuzzing techniques [31], I injected a wide range of unexpected and malformed inputs into network-related system calls to assess their response patterns.

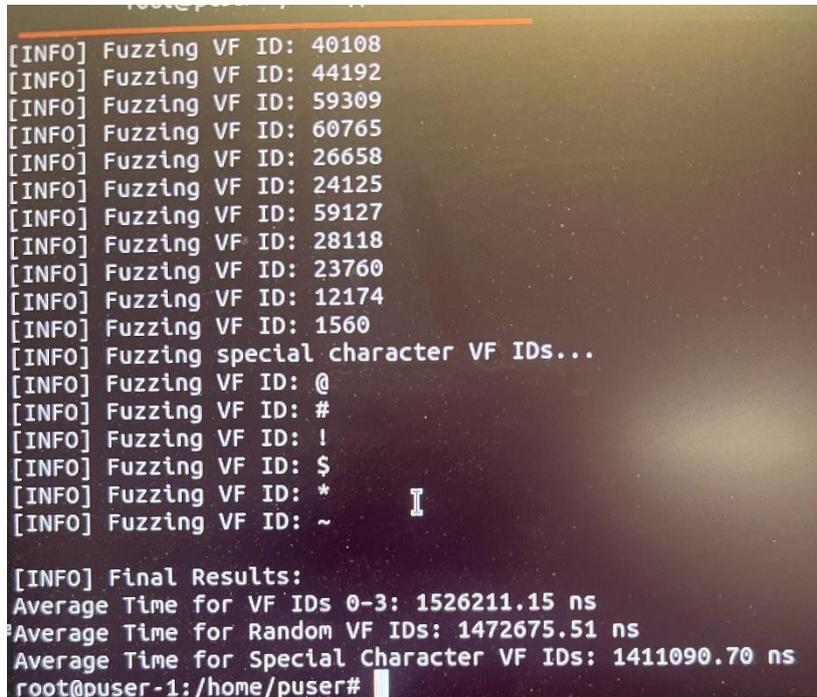

Fig 5.1 - Hash Table access timing test results screenshot

These results were intriguing as there was a measurable timing difference between Occupied, Unoccupied, and special charter queries, this led to the research question:

- **Is it possible to infer if a VF ID is occupied based on timing measurements?**

To begin researching this question, a series of methods were created to see if this was possible for an unprivileged user to infer if a VF ID is occupied **not** based on time via the PF (aka on the hypervisor not within a VM) [4], [13]. It was found that the most revealing system utility in this context is `readlink()`, which can be used to query VF IDs in /sys/class/net/.

When executed against a valid VF, `readlink()` returns an address corresponding to the physical device. Conversely, when querying an unoccupied or non-existent VF ID, the command returns nothing. This behaviour provides a clear method for determining which VFs are currently in use. Shows that this can be inferred without time which lowers the impact.

Fig 5.2 - hash table query using `readlink()` screenshot

Additional testing was conducted using the " `ip addr show` " bash script, which lists network interfaces and their associated parameters, although this method does not explicitly indicate VF presence in all cases, timing variations in execution provided a 63% success rate in distinguishing active from inactive VFs. The utility's reliance on kernel interactions introduces slight delays when processing different types of network devices, creating an exploitable timing signal. This test somewhat proves that it's possible to denote if a VF ID is occupied.

*This test, takes the known occupied IDs (0-3) adds them together, and divides them by 4, it also takes a random sample of 4 unoccupied VF IDs from 4- 10 and does the same, this was a regression test ran 1000 times to see how many times (out of 1000) the average access time for occupied was higher than unoccupied, this proved to be the most conclusive test using "`ip addr show (x)`"*

Fig 5.3: Hash table query results using "ip a" and timing analysis

Another user-space approach involved is leveraging " `lspci` ", which interacts with the PCI bus to enumerate available network devices. Although `lspci` directly queries the PCI bus to list devices, independent of whether drivers are loaded [3], [4]. It accesses the PCI configuration space via interfaces like /sys/bus/pci/devices/, bypassing device-specific drivers. This allows it to detect all PCI devices, even those without associated drivers [4], [37].

This command exhibited the highest variance in execution time, while `lspci` is a more indirect method for detecting VFs, the timing fluctuations observed reinforce the potential for side-channel inference. This test proves that VF id occupancy can be denoted using time in user space to an unprivileged user, although this is deemed out of scope as the ICE driver is bypassed.

*This test calculates the same averages, but unlike printing out which was higher upon every run, it just provided the overall average of occupied access times v unoccupied. As the difference for this in user space "`lcpci`" is the least prone to noise.*

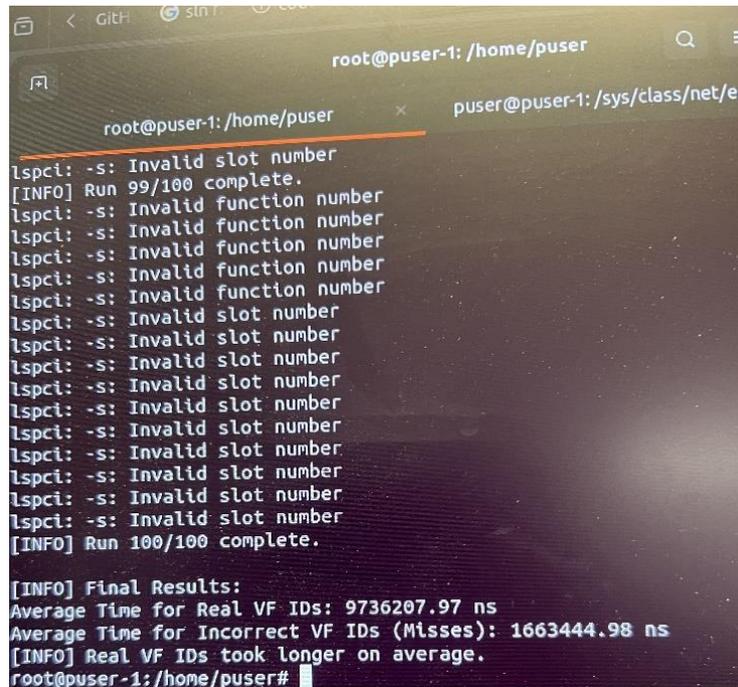

Fig 5.4 - Hash Table Query using "`lspci`" and timing analysis.

The results of these user-space tests indicate that while explicit enumeration methods such as `readlink()` reveal VF presence directly, even commands that do not provide a clear binary response can still leak information through execution time variations as results show a 63% success rate in distinguishing occupied from unoccupied VFs using `'ip addr show'`, and higher accuracy when using `'lspci'`. This confirms that VF presence can be inferred via timing analysis. These findings align with *"kernelsnitch"* proving that timing-based side-channel vulnerabilities often emerge not from explicit leaks but from subtle execution time differences in kernel operations [31].

5.3 Kernel Instrumentation for Hash Table Timing Analysis

While user-space inference attacks revealed timing variations, deeper analysis was needed to validate the user-space findings to confirm if it's possible to infer if a VF ID is occupied based on timing measurements.

To achieve this, kernel instrumentation was added to " `ice_get_vf_by_id()` "within `ice_vf_lib.c` in the `/src` folder of the ICE Driver, this is the function responsible for querying VF entries [4] RTDSC (Read Time Stamp Counter) was used as a mechanism to gather accurate timing measurements [38].

This function maps Virtual Function (VF) IDs to their corresponding structures using a hash table with Read-Copy Update (RCU) locks to ensure concurrency [4], [33]. The instrumented version introduces RTDSC timestamp counters to measure lookup times without altering functionality [38]. The key modifications include:

- Looping through VF ID queries multiple times for statistical accuracy.
- Recording timestamps before and after lookups to measure execution duration.
- Logging times for successful and failed lookups, highlighting timing discrepancies.

The function iterates over a defined range of VF IDs, capturing timestamps before validation. If a VF ID is invalid (U16_MAX), it logs the failure and moves to the next iteration. For valid IDs, it performs an RCU-protected hash table lookup [33], [43]. If the VF exists, " `kref_get_unless_zero()` "ensures validity before recording the final timestamp [4]. Failed lookups are also timed and logged, enabling direct comparison between occupied and unoccupied VF IDs.

5.3.1 Impact of Instrumentation

This instrumentation does not alter the ICE driver's behaviour, as it passively measures execution times without affecting control flow. The analysis confirmed that occupied VF IDs consistently returned faster than unoccupied ones. Additionally, hash collisions increased lookup times, further amplifying side-channel leakage risks.

Cross-referencing these findings with user-space timing measurements validated that VF presence, hash table occupancy, and induced collisions could all be inferred [43]. To mitigate these risks, constant-time lookups and randomized hash functions should be considered, reducing the potential for timing-based inference attacks in virtualized environments.

Fig 5.5 - Intel Ice Driver `src/ice_vf_lib.c` instrumented code vs standard code (proof no extra functionality was added)

Instrumentation was introduced at multiple points within "`ice_get_vf_by_id()`" to capture timing variations under different conditions [4]. The results showed a clear distinction between occupied and unoccupied VF lookups, with occupied IDs **returning results with 10 times more clock cycles on average** than unoccupied IDs [31]. This discrepancy arises from the fact that successful lookups require more computational steps compared to failed searches, [31], [43].

Fig 5.6 - post-Instrumentation hash table query timing analysis results

## 5.4 Summary of Findings

The results of this study demonstrate that timing-based inference attacks on the ICE driver's hash table are feasible in multi-tenant environments [31], [43]. By leveraging a combination of explicit enumeration techniques (`readlink() & ip addr show`) and timing-based inference, an attacker can determine VF presence with a high degree of confidence. Furthermore, controlled hash collisions can be used to amplify side-channel leakage or degrade performance, making this an exploitable attack vector.

### 5.4.1 Impact of Findings

The findings highlight the importance of constant-time hash table lookups and randomized hash functions as potential countermeasures to mitigate these risks [43]. Without such defences, adversaries in shared cloud environments could extract sensitive metadata about other tenants, posing a significant security threat [12], [31].

## 5.5 Potential Attack Scenario
A potential attack scenario, based on these results would affect cloud data centers for example, and CDC will be the presumed Host for the attack.

Host Prerequisites:
A host to be running Linux, with an Intel NIC, multiple tenants across VFs that are designed to be isolated from each other.

Attacker: Has unprivileged access to the system but is allowed to create VF ids.

Attack logs on and uses the `lspci` and RTDSC to measure and compare access time for hits and misses when querying the VF ID, from this they denote what VF IDs are occupied or not.

It is at this stage the attacker has multiple options:

- Create a targeted Dos Attack based on the VF ID they now know is occupied.
- Create a targeted Dos Attack based on a busy VF ID, that displays a higher return time than an unbusy occupied VF ID.
- Create More VF IDs to intentionally cause collisions with known to be occupied IDs.
- Redirect traffic from Busy VM to own via ring descriptor manipulation.

# Chapter 6: Exploiting RCU-Based VF Management to Induce OOM and Device Failures

This chapter examines a bug in SR-IOV Virtual Function (VF) management, where rapid creation and deletion of VFs leads to either:

- Kernel log flooding and device resets due to failed VF allocation attempts.
- Out-of-Memory (OOM) conditions, resulting in a system-wide crash and black screen [39].

This investigation stems from what was thought to be a failed test / inconclusive result from chapter 5:

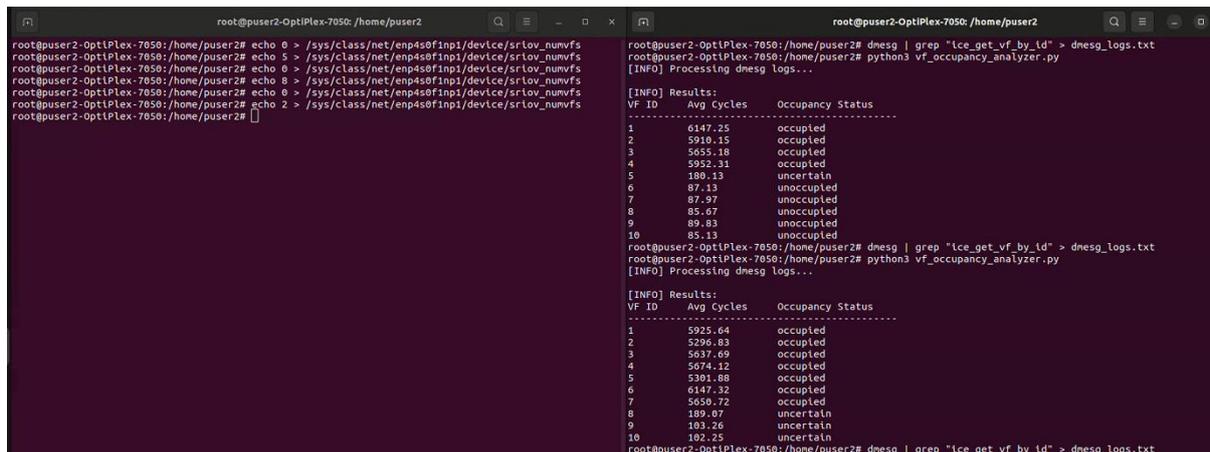

Fig 6.1 - Timing results screenshot that suggests a stale pointer is present.

In the above image, a hash table reset is observed, and I attempted to replace the existing 8 Virtual Functions (VFs) by writing 2 new VFs after the reset [4], [31]. However, on the left side of the image, the occupied VF IDs still display outdated ("stale") values, even though they should have been updated. This inconsistency led to the formulation of the research question.

**How does the RCU mechanism contribute to stale data persistence in SR-IOV VF management, and what are the underlying causes of delayed entry invalidation?**

The root cause of this behaviour lies in the lack of explicit RCU (Read-Copy-Update) synchronization mechanisms, such as `call_rcu()` and `synchronize_rcu()`, within the ICE driver's VF management system [4], [33], [42]. RCU enables concurrent read access while deferring memory reclamation until all pre-existing readers have exited their critical sections [33]. However, in the ICE driver, the absence of explicit RCU synchronization results in stale VF entries persisting longer than expected after deletion [42].

When Virtual Functions (VFs) are freed, the function `ice_free_vf_entries()` in *ice_sriov.c* performs deletion from the hash table using `hash_del_rcu()` [4], [43]. The function iterates over the hash table, removes each VF entry using `hash_del_rcu()`, and subsequently calls `ice_put_vf(vf)`, which decrements the reference count of the VF [4]. However, `hash_del_rcu()` only removes the VF entry from the hash table logically but does not guarantee immediate memory reclamation [4], [33]. The ICE driver does not call `synchronize_rcu()` after removing VFs, which means that existing RCU readers may continue accessing stale VF data even after deletion [4], [33]. Explicitly calling `synchronize_rcu()` after freeing the VFs would ensure that all pre-existing RCU readers exit before the deletion completes, preventing inconsistencies in VF state management [33], [42].

In `ice_put_vf()`, which is defined in `ice_vf_lib.c`, the ICE driver manages reference counting for VF structures [4]. Within `ice_release_vf()`, the function `vf->vf_ops->free(vf)` is responsible for freeing the VF memory [4]. However, this operation is performed without using *call_rcu()*, meaning RCU readers may still be accessing the now-freed memory [4], [33]. If a lookup function such as `ice_get_vf_by_id()`, defined in `ice_vf_lib.c`, is executed immediately after a reset, it may return a stale pointer [42]. The function `ice_get_vf_by_id()` acquires the RCU read lock, iterates through the hash table, and returns a VF entry if its reference count is nonzero [4], [33]. Since `call_rcu()` is not used in `ice_release_vf()`, the memory for the VF structure may be reclaimed while it is still accessible through `ice_get_vf_by_id()`, leading to potential use-after-free conditions [33], [40]. [42].

The absence of `call_rcu()` and `synchronize_rcu()` has multiple implications. Stale VF entries remain accessible after their intended deletion, leading to inconsistent VF states. The potential for use-after-free conditions increases, as `ice_get_vf_by_id()` may return references to freed memory [40]. Memory fragmentation may also occur, as objects remain allocated longer than necessary. Implementing `call_rcu()` within `ice_release_vf()` would ensure that memory is not reclaimed until all active RCU readers have finished accessing it [33]. Similarly, incorporating `synchronize_rcu()` in `ice_free_vf_entries()` would prevent stale VF data from persisting in the hash table beyond its expected lifetime [42], [43]. Addressing these issues would enhance the reliability and efficiency of the ICE driver's VF management system.
[33].

The diagram illustrates how improper RCU synchronization in the ICE driver could lead to stale VF pointers, potential Use-After-Free (UAF) conditions, and Out-of-Memory (OOM) errors [39], [40]. VFs are allocated via `ice_alloc_vfs()` and added to a hash table with `hash_add_rcu()`. When freed using `ice_free_vf_entries()`, `hash_del_rcu()` removes them logically but does not reclaim memory immediately. Stale pointers persist in `ice_get_vf_by_id()`, leading to UAF [40]. Rapid VF creation and deletion of fragment memory, filling SLAB caches and preventing proper reuse. Eventually, the kernel exhausts memory, triggering the OOM killer. This highlights security and stability risks due to missing `synchronize_rcu()` and `call_rcu()`.

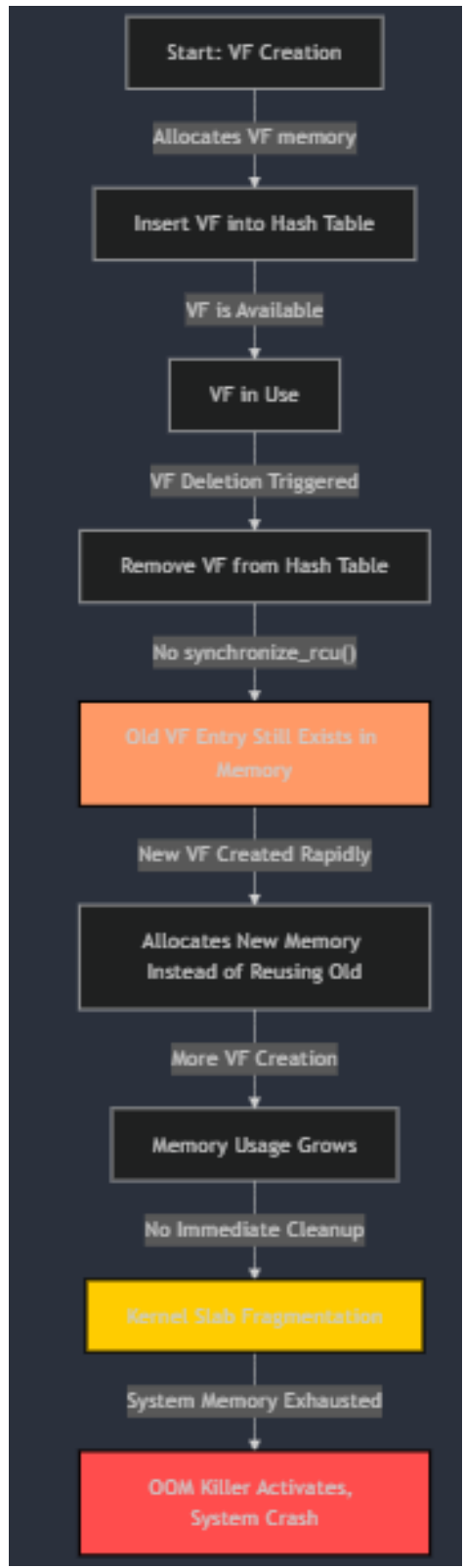

Fig 6.2 - Flowchart of possible Denial of Service weakness that stems from improper RCU implementation.

## 6.1 Experimental Setup

The following tests were conducted on an SR-IOV-enabled system using an Intel NIC with hardware virtualization support. Two different command sequences were executed to observe distinct failure conditions:

Test Case, Command, Expected Behaviour, Actual Outcome

VF Creation Spam:

```
while true; do echo 4 > /sys/class/net/<Device-Name>/device/sriov_numvfs; done
```

repeatedly create VFs without deletions, VF allocation failures, `dmesg` errors, device resets

Parallel VF Create/Delete Flood:

```
for ((;;)); do echo 0 > /sys/class/net/<Device-Name>/device/sriov_numvfs & echo 4 > /sys/class/net/<Device-Name>/device/sriov_numvfs & done
```

Overload RCU by spamming create/delete cycles, Memory exhaustion, OOM Killer activation, system crash (black screen)

## 6.2 VF Creation Spam – Kernel Log Errors

"`while true; do echo 4 > /sys/class/net/<Device-Name>/device/sriov_numvfs; done`"

Observed Behaviour:

Repeated VF creation attempts to cause device resets and failed resource allocation messages in `dmesg`.

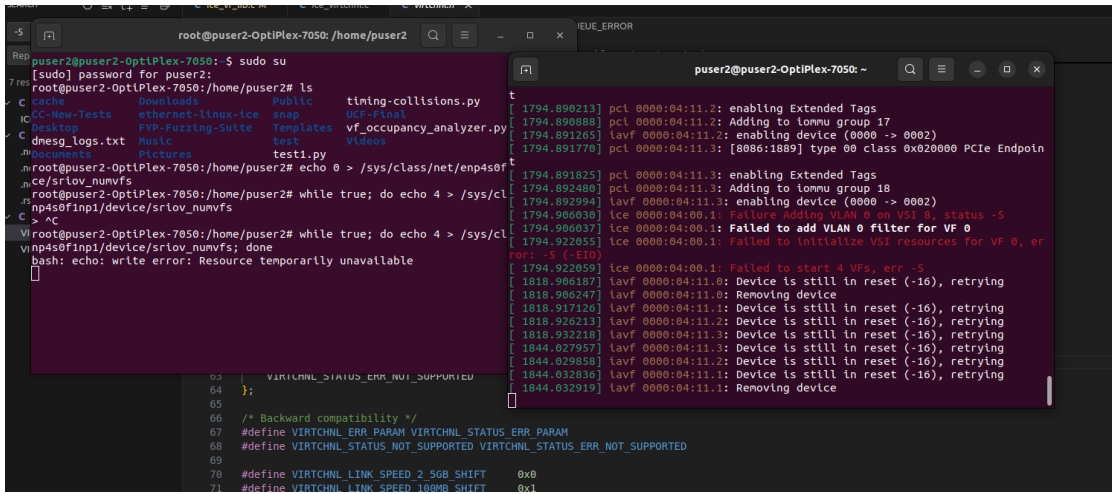

Fig 6.3 - Hash table Creation Spam results Screenshot.

Kernel logs show errors like:

- `Failed to add VLAN 0 filter for VF 0`
- `Failed to start 4 VFs, err -5`
- `Device is still in reset (-16), retrying`

The system remains operational, but logs are flooded, potentially degrading performance.

Root Cause:

1. The command attempts to create VFs indefinitely, exceeding hardware or PCI resource limits.
2. The driver fails to allocate more VFs but does not exhaust system memory.
3. The kernel retries allocation instead of immediately failing, resulting in continuous `dmesg` errors.

## 6.3 Parallel VF Create/Delete Flood – OOM Crash
Command:

```
" for ((;;)); do echo 0 > /sys/class/net/<Device-Name>/device/sriov_numvfs
& echo 4 > /sys/class/net/<Device-Name>/device/sriov_numvfs & done "
```

Observed Behaviour:

The system locks up and crashes, displaying a black screen.

The OOM Killer terminates multiple processes, including bash, systemd, and network services.

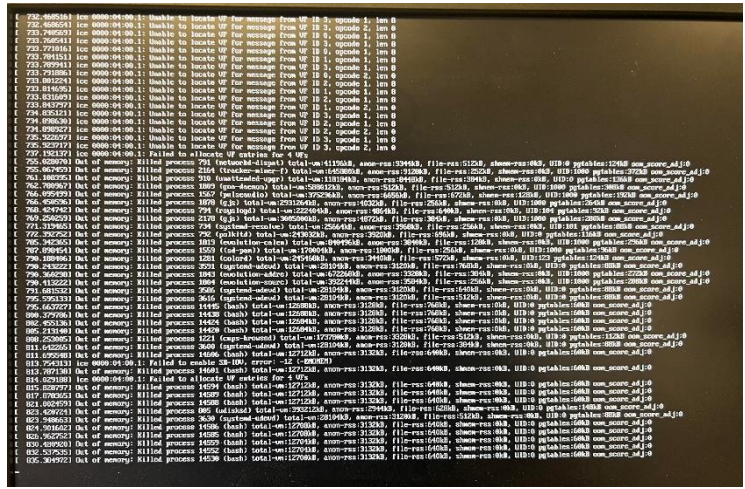

Fig 6.4 - OOM Killer activated Screenshot

dmesg logs before the crash indicate memory exhaustion:

The system becomes unresponsive and requires a hard reboot.

Root Cause:

- This command runs parallel background processes (&) for VF creation and deletion at high speed.
- The RCU grace period cannot be completed fast enough, leading to deferred VF deletions piling up in memory.
- Old VF structures remain allocated, consuming kernel memory until it is completely exhausted.
- The OOM Killer forcefully terminates processes, but if critical services are killed, the entire system crashes.

## 6.3 Exploit Analysis: How RCU Hash Tables Enable This Attack

The root cause lies in RCU (Read-Copy-Update) hash tables, which manage the VF lifecycle without locks. The attack takes advantage of RCU's three key behaviours:

RCU Mechanism & How the Attack Exploits It:

- Lockless Reads, Reader's access data without locks, Rapid VF creation increases concurrent readers, forcing stale VF entries to persist [33]
- Copy-on-Write Updates, New data is written while old data waits for deletion, Continuous VF creation generates tons of old, unfreed copies
- Deferred Reclamation, Old data is only freed after all readers finish, VF churn delays cleanup indefinitely, leading to memory exhaustion

- By manipulating timing and concurrency, the attack prevents RCU from cleaning up stale objects, leading to unbounded memory growth and system failure.

The absence of `synchronize_rcu()` in Virtual Function (VF) deletion results in delayed memory reclamation, leading to stale hash table entries persisting beyond their expected lifetime. As highlighted in [42], deferred memory management in procrastination-based synchronization introduces unexpected memory fragmentation, increasing the likelihood of Out-of-Memory (OOM) conditions under rapid allocation and deallocation cycles [42]. Timing analysis of the ICE driver demonstrates that, after VF deletions, hash table lookups briefly return incorrect timing values, corresponding to previously occupied VF entries, indicating a use-after-free risk [4]. Additionally, when VFs are continuously created and destroyed in a high-frequency loop, kernel memory consumption escalates beyond system limits, ultimately triggering an OOM crash.

## 6.4 Lack of synchronize_rcu() and Its Role in SR-IOV VF Vulnerabilities

This section extends the previous analysis by focusing on the role of `synchronize_rcu()` in mitigating memory exhaustion and use-after-free (UAF) conditions in SR-IOV Virtual Function (VF) management.

### 6.4.1 The Role of synchronize_rcu() in RCU Memory Management

RCU (Read-Copy-Update) is a synchronization mechanism that allows lockless read operations, enabling high-performance lookups in data structures such as hash tables. However, RCU defers memory reclamation, meaning that deleted objects remain accessible until all readers exit their critical sections. This behaviour is efficient under normal operation, but when rapid VF creation and deletion cycles occur, it leads to:

- Memory exhaustion: Old VF structures accumulate, consuming kernel memory indefinitely.
- Use-after-free (UAF): Readers may access freed VF structures before they are properly reclaimed.

The function `synchronize_rcu()` ensures that all ongoing RCU readers finish before freeing memory. The absence of `synchronize_rcu()` in the SR-IOV VF handling code allows stale VF objects to persist longer than necessary, leading to system instability.

```
/**
 * ice_release_vf - Release VF associated with a refcount
 * @ref: the kref decremented to zero
 *
 * Callback function for kref_put to release a VF once its reference count has
 * hit zero.
 */
static void ice_release_vf(struct kref *ref)
{
    struct ice_vf *vf = container_of(ref, struct ice_vf, refcnt);

    if (vf->migration_active)
        ice_migration_uninit_vf(vf);

    vf->vf_ops->free(vf); // memory is freed, but RCU readers may still access it!
}
```

Fig 6.5 - Screenshot of affected function `ice_release_vf` within *ice_vf_lib.c*

Why Is This a Problem?

1. `ice_put_vf()` decreases the reference count.
2. Once the reference count hits zero, `ice_release_vf()` is called.
3. `ice_release_vf()` immediately frees the vf structure without ensuring that RCU readers have finished.
4. Since RCU queries (such as `ice_get_vf_by_id()`) can still reference the VF, stale pointers may be accessed, leading to UAF.

Mitigation:

```c
static void ice_release_vf(struct kref *ref)
{
    struct ice_vf *vf = container_of(ref, struct ice_vf, refcnt);

    if (vf->migration_active)
        ice_migration_uninit_vf(vf);

    synchronize_rcu();  // ensures no active readers before freeing
    vf->vf_ops->free(vf); // memory is freed, but RCU readers may still access it!
}
```

Fig 6.6 - proposed mitigation for affected function `ice_release_vf` withn *ice_vf_lib.c*

## 6.5 Publication

The Novel findings from this chapter have been submitted to ISSC 2025, see Appendix A for the submission paper.

## 6.6 Disclosure and Industry Response

Intel reviewed the reported OOM and RCU synchronization issue in the ICE driver and concluded it is not a security vulnerability, as only privileged (root) users can trigger the behaviour. While no direct security impact was identified, the findings remain relevant for system stability and reliability, particularly in high-load environments, achieving informative status with a validity score of 100%.

# Chapter 7: Conclusions and Future Work

## Conclusions

This project has provided a thorough security assessment of the Intel ICE driver for the E810 Ethernet Controller, leveraging fuzzing techniques, static analysis, and timing-based side-channel investigations. The primary objective was to evaluate the resilience of the driver against malformed inputs and to identify potential security risks that could be exploited in enterprise and data center environments. The results indicate that while the ICE driver implements robust input validation mechanisms, preventing common memory corruption vulnerabilities, there are notable concerns regarding timing-based side-channel leaks and the handling of virtual functions (VFs) in a multi-tenant environment.

The fuzzing experiments, which tested various driver interfaces including the Admin Queue, debugfs, and user-space configuration tools, did not reveal critical memory corruption flaws. This suggests that the ICE driver has been developed with strong input validation and error handling mechanisms, reducing the risk of traditional buffer overflows, format string vulnerabilities, and command injection attacks. The debugfs and Admin Queue fuzzing results confirmed that malformed inputs were properly handled, ensuring that unintended crashes, privilege escalations, or data leaks were not observed.

However, despite its robustness against direct input attacks, the ICE driver exhibited vulnerabilities related to side-channel inference. The study demonstrated that an unprivileged user could infer the presence and activity of VFs by analysing execution time discrepancies in hash table lookups. This is particularly concerning in shared cloud environments where isolation between virtualized tenants is crucial. The findings showed that occupied and unoccupied VF IDs exhibited measurable timing differences, potentially allowing an attacker to map the network environment and target specific VFs for denial-of-service (DoS) attacks or traffic manipulation.

Further kernel instrumentation confirmed that hash table lookup times varied based on occupancy, reinforcing the need for constant-time operations and improved hash function design to mitigate these risks as another significant discovery was related to the ICE driver's handling of VF creation and deletion in an SR-IOV-enabled environment.

Through controlled testing, it was observed that rapid allocation and deallocation of VFs could lead to resource exhaustion, kernel instability, and system-wide crashes. The root cause was traced to the lack of explicit synchronization mechanisms, such as `synchronize_rcu()`, which resulted in stale VF entries persisting longer than expected. This behaviour created the potential for use-after-free conditions and Out-of-Memory (OOM) scenarios, highlighting the necessity for improved memory management and synchronization within the ICE driver.

# Appendices

## Appendix A: List Of Publications:

ISSC Submission

# Identifying Linux Kernel Instability due to Poor RCU Synchronization

Oisin O'Sullivan
Department of Electronic & Computer Engineering
University of Limerick
Limerick, Ireland.
21304971@studentmail.ul.ie

Colin Flanagan
Department of Electronic & Computer Engineering
University of Limerick
Limerick, Ireland.
colin.flanagan@ul.ie

Eoin O'Connell
Department of Electronic & Computer Engineering
University of Limerick
Limerick, Ireland.
eoin.oconnell@ul.ie

*Abstract*—Read-Copy-Update (RCU) is widely used in the Linux kernel to manage concurrent access to shared data structures. However, improper synchronization when removing RCU-protected hash table entries can lead to stale pointers, inconsistent lookups, and critical use-after-free (UAF) vulnerabilities. This paper investigates a driver-level synchronization issue arising from the omission of explicit `synchronize_rcu()` calls during hash table updates, using a discovered weakness in the Intel® ICE network driver's Virtual Function (VF) management. Previous kernel vulnerabilities, such as a bug in the Reliable Datagram Sockets (RDS) subsystem, show how improper RCU synchronization can directly cause kernel crashes. Experimental results demonstrate that removing VF entries without proper synchronization leaves transient stale entries, delays memory reclamation, and results in significant memory fragmentation under rapid insert/delete workloads. Broadly speaking, RCU hash tables are widely used across many Linux kernel subsystems, including networking, virtualization, and file systems. Improper synchronization can lead to severe memory fragmentation, kernel instability, and eventual Out-of-Memory (OOM) conditions. Mitigations are proposed, recommending explicit insertion of `synchronize_rcu()` calls to ensure timely and safe memory reclamation. These findings reinforce established best practices for RCU synchronization, highlighting their importance for maintaining kernel stability and memory safety.

*Keywords— RCU, kernel synchronization, hash tables, ICE driver, memory fragmentation, use-after-free*

## I. INTRODUCTION

Modern operating systems and many applications frequently employ lock-free data structures to achieve high concurrency [1]. The Linux kernel's RCU (Read-Copy-Update) is a widely used mechanism which allows readers to access data without locks while writers defer freeing or updating data until no readers are using it [1], [2]. A typical pattern is to remove an element from an RCU-protected list or hash table using `call_rcu` or similar deferred freeing, or by explicitly waiting for an RCU grace period via `synchronize_rcu()` [1]. Failing to synchronize properly after deletion can leave stale entries accessible to readers, risking inconsistency and use-after-free (UAF) errors [1] - [4]. For instance, a bug in the RDS network subsystem was caused by freeing a socket immediately after removal from an RCU hash table, readers could still find the freed socket, leading to a UAF reported by "*syzkaller*" [4], [5], [6]. The fix involved deferring the freeing of memory until after an RCU grace period [4]. Similarly, in the eBPF subsystem [7], a lack of RCU grace in freeing inner map objects led to potential UAF [8], which was fixed by invoking deferred freeing (`via call_rcu()`) to ensure memory wasn't reclaimed until all readers were done [3].

These examples illustrate that RCU misuse can corrupt data or crash the kernel, motivating careful handling of object lifecycles.

This work focuses on the Intel® ICE Ethernet driver as a practical test case to explore the impact of missing `synchronize_rcu()` in hash table management [9]. The ICE driver maintains a hash table of VF (Virtual Function) metadata structures for SR-IOV virtualisation [10]. RCU protects each VF entry for lockless lookups [1], [9]. When VFs are removed, for example, when an administrator disables some VFs or during PCIe VF teardown, the driver code deletes the VF entries from the hash table [9]. In the studied ICE implementation, these deletions use `hash_del_rcu()` to remove the entry, and then drop the VF reference count, which leads to freeing the VF structure immediately if no other references remain [8]. Notably, no `synchronize_rcu()` or similar barrier is called after removing the entries [1], [2]. This means the driver relies on RCU list-del semantics and reference counting to avoid UAF [2], [9]. However, the absence of an explicit RCU sync can potentially leave a window where other CPU cores might still hold references to the deleted VF or see it via RCU traversal [11]. This work examines the consequences of omitting `synchronize_rcu()` calls when deleting entries from RCU-protected hash tables, using the discovered weakness in ICE driver as a practical use case for experiments [3], [5].

## II. METHODOLOGY

To investigate the effects of failing to include `synchronize_rcu()` in hash table deletions, experiments were designed around the Intel® ICE driver's VF management routines [8], [9]. The methodology involved both targeted stress tests to evaluate memory usage and system stability under rapid VF churn [12].

All tests were performed on an Intel® e810 controller which supports SR-IOV, an Intel® Core i7-7700 (Kaby Lake) processor, running Linux Kernel version 6.8.0 with the ICE driver version 1.16.3 installed. [1], [9], [10].

VF Creation/Deletion Test Loop: A command was crafted to create and destroy VFs on the ICE NIC in a loop simultaneously. Creation and deletion were done via standard sysfs interfaces using the following bash script, which enables N VFs, then writes 0 to rapidly disable them [9].

```
for ((;;)); do echo 0 > /sys/class/net/<Device-Name>/device/sriov_numvfs & echo N >
```

```
/sys/class/net/<Device-Name>/device/sriov_numvfs &
done
```

In each cycle, the number of VFs is varied (up to the device's maximum, 64 in this case) and the interval between create and destroy operations [9], [13]. By running create/destroy iterations in a tight loop, the aim is to force the driver to exercise its VF allocation and teardown logic rapidly [9], [13]. This is an extreme scenario akin to live migration [14], [15]. VFs are not typically toggled this quickly, but it's useful to reveal any race conditions or accumulation of deferred frees [12].

Memory Usage and OOM Monitoring: During the stress loops, kernel memory usage will be tracked continuously via `/proc/meminfo` and `dmesg` logs for any OOM killer activity [16], [17], [18]. The system's OOM behaviour was set to panic on OOM for a clear signal. Two possible outcomes were anticipated under extreme churn, either a gradual memory buildup leading to an OOM trigger, or a hard crash if a UAF corrupted memory [3], [4], [12], [16].

RCU Grace period timing analysis: To assess the impact of missing `synchronize_rcu()` in the ICE driver's VF deletion process, tests were designed to focus on timing discrepancies in hash table lookups. By executing VF creation and deletion, we measured how long stale VF entries persisted before memory was reclaimed. Using a "KernelSnitch" inspired timing analysis test [19], timestamps were logged when a VF entry was removed and when its memory was freed.

This test methodology extends beyond the Intel® ICE driver and applies to other RCU-protected kernel subsystems [20]. RCU-based hash tables are widely implemented in networking, storage, and security modules, making it essential to evaluate synchronization robustness systematically [20], [21]. By combining these methods, timing measurements, limit testing, and failure testing, A comprehensive test plan is created [3], [4], [9], [12], [19]. Regression Tests were conducted to ensure reproducibility of any anomalies [23]. In all cases, tests were conducted on an isolated test system and with root privileges (since SR-IOV and debug interfaces require it) [10], [13]. This methodology aims to capture a brief stale pointer existence and/or memory exhaustion attributable to missing RCU synchronization [3], [4], [12].

## III. RESULTS

The most notable result came from the stress test of repeatedly enabling/disabling VFs. It was found that without any explicit throttling or RCU sync, continuous VF churn led to steadily increasing memory usage by the kernel, culminating in an OOM condition [12], [16]. After creating 64 VFs and immediately deleting them, the test system's free memory plummeted and the OOM killer engaged [12], [16]. The `dmesg` logs showed multiple allocation failures in the ice driver and ultimately an OOM kill targeting either our test process or other processes [12]. In one run, the observed system messages were as follows:

```
ice_alloc_vf_res:   allocation   failure,   order:3,
mode: GFP_KERNEL" followed by "Out of memory: Killed
process 1234 (modprobe) total-vm:…
```

This indicates that the driver failed to allocate a contiguous block (order 3 is 8 contiguous pages) for a VF resource, likely due to fragmentation [24], and the overall memory was exhausted enough to invoke OOM killer [12], [16].

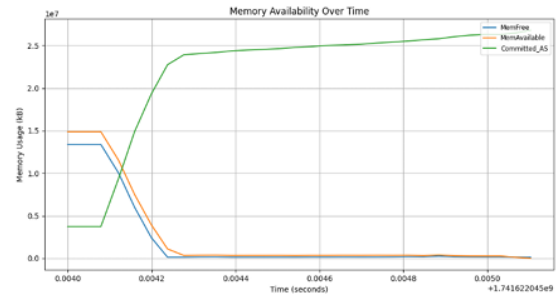

Fig.1

As seen in the above figure, when the OOM killer ran, the system still reported an amount of memory "available" in free or `/proc/meminfo` [18], in this case, about 120 MB of RAM was still free when the OOM occurred [12], [16], [17].

This seemingly paradoxical situation is explained by memory fragmentation, some free memory existed, but not in sufficiently large contiguous chunks to satisfy certain allocation requests [24]. The kernel's allocator, unable to service a high-order allocation for the NIC driver, eventually gave up and triggered OOM despite plenty of scattered free pages [20], [21]. This has been documented by others, a user on an ARM system saw OOM kills also with ~120MB free, and only by manually compacting memory could they prevent the OOM [25]. Our scenario is analogous, as each VF creation allocates various structures. Rapidly allocate-free cycles without full synchronization aggravated fragmentation and memory strain [12].

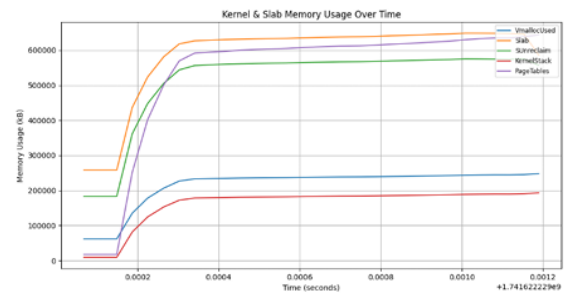

Fig. 2

To quantify memory growth, the kernel and slab memory usage were logged. As shown in Fig 2, Slab, SUnreclaim, and PageTables memory rapidly increased and stabilized at elevated levels, even after VF deletion, suggesting memory fragmentation and delayed reclamation [12], [28], [24]. The persistent high slab usage indicates inefficient memory freeing, as frequent updates deferred RCU grace periods, causing kernel objects to linger [18]. This confirms our hypothesis: rapid VF churn without proper synchronization exacerbates fragmentation, eventually triggering allocation failures and OOM conditions, despite some memory appearing free [12], [16].

Despite VF deletions, Slab usage remained high, indicating delayed memory freeing. This suggests that frequent updates deferred RCU grace periods, causing objects to persist longer than expected [12]. Ultimately, rapid VF cycling led to persistent memory growth, confirming that fragmentation and delayed reclamation contributed to potential OOM scenarios [12], [16]. It's worth noting that the OOM condition is exacerbated by the fragmentation issue mentioned [16], [25]. When the OOM killer triggered, it wasn't that the system had zero free pages, just not the right kind or grouping [16]. Logs that the buddy allocator was failing to find an order-3 or order-4 page were also observed. The continuous allocate-free churn of large objects leads to fragmentation where free memory is in small pieces [12], [16]. The memory compaction daemon (`kswapd/kcompactd`) was not keeping up because our workload was constantly consuming and releasing memory [26]. Essentially, the driver was requesting memory in a pattern that the kernel struggled to fulfill after sufficient fragmentation, causing allocation failures that cascaded into OOM [9], [12].

Additionally, during stress testing involving rapid VF creation and deletion cycles without explicit synchronization, occasional system instability consistent with documented use-after-free (UAF) vulnerabilities was observed [3], [4], [12], [27]. Specifically, the terminal window monitoring the process repeatedly froze and terminated unexpectedly, strongly indicative of memory corruption due to stale pointers accessing already-free memory [8]. Such instability aligns closely with findings from [27], highlighting that inadequate synchronization in high-frequency allocation and deallocation scenarios exacerbates conditions leading to UAF bugs [8].

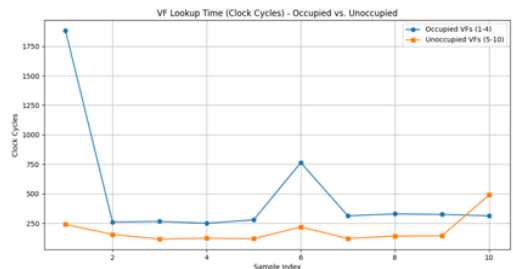

Fig. 3

The results from the RCU grace period timing analysis indicate that timing values associated with occupied VF entries persist for a short period after deletion, suggesting that memory is not immediately reclaimed [12], [19]. Figure 3 illustrates that lookup times for deleted VFs remained consistent with occupied VF entries for a short duration before eventual memory release. This confirms that the lack of `synchronize_rcu()` delays proper cleanup, causing stale data to persist longer than expected [12].

In summary, stress testing of rapid VF creation and deletion without explicit RCU synchronization led to steadily increasing kernel memory usage, culminating in OOM conditions [12]. Despite `/proc/meminfo` reporting available free memory, fragmentation prevented contiguous allocations, forcing the OOM killer to terminate processes [16], [17], [24]. Timing analysis (Fig. 3) revealed that stale VF timing values persisted as if they were occupied for a short period after deletion, inferring delayed memory reclamation due to missing `synchronize_rcu()`. Kernel memory tracking (Fig. 2) showed Slab, SUnreclaim, and PageTables usage remained elevated, reinforcing inefficient memory freeing and RCU grace period delays [12], [18]. Additionally, occasional system instability suggested potential UAF risks caused by stale pointers referencing freed memory [8], [27]. While memory was eventually reclaimed, it was not fast enough to prevent crashes, confirming that fragmentation and delayed cleanup exacerbated OOM failures [12].

IV. DISCUSSION

Our findings demonstrate a critical trade-off in kernel synchronization: asynchronous reclamation versus explicit synchronous waits [28]. The ICE driver chose asynchronous reclamation, freeing memory immediately after deletion from RCU-protected hash tables without invoking `synchronize_rcu()` [9], [12]. While this approach avoids immediate blocking, it introduces potential for stale references and memory instability, as evidenced by our results.

While adding `synchronize_rcu()` ensures proper memory reclamation, alternative strategies may provide performance-efficient solutions [4]. One such approach is using `call_rcu()`, which defers cleanup until existing RCU readers complete their critical sections without forcing immediate execution [1], [4]. Another method involves rate-limiting VF churn, preventing excessive creation and destruction cycles within short timeframes, thereby reducing fragmentation risk. Intel® may have avoided using synchronize_rcu() due to concerns about its impact on throughput in latency-sensitive workloads. However, as seen in other kernel components, hybrid techniques that balance deferred freeing with controlled allocation policies can mitigate both performance overhead and memory exhaustion [6], [7].

These observed issues align closely with previously documented kernel vulnerabilities involving use-after-free (UAF) conditions [3], [4]. Specifically, our experiments demonstrated occasional system instability, with terminal windows freezing and terminating unexpectedly during rapid VF creation and deletion cycles [8], [12]. This behavior strongly indicates memory corruption due to stale references accessing already freed memory, matching the characteristics of UAF vulnerabilities outlined in prior research [3], [4], [8], [27].

Moreover, our stress tests revealed that rapid allocation and deallocation cycles without synchronization caused severe memory fragmentation, ultimately triggering out-of-memory (OOM) conditions despite the presence of sufficient total free memory [12]. [16] , [24]. This paradoxical scenario occurs because fragmented memory lacks sufficiently large contiguous blocks required for certain kernel allocations. Similar phenomena have been documented previously, further confirming the risks of inadequate synchronization [12], [16], [24].

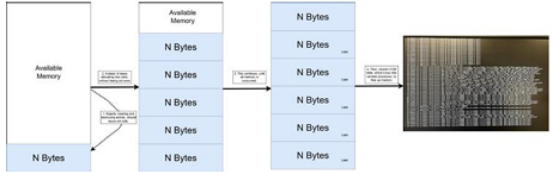

Fig. 4

To mitigate these issues, introducing explicit synchronization calls such as `synchronize_rcu()` during VF teardown is recommended [1], [9]. This ensures memory is only freed after all concurrent readers have exited their critical sections, significantly reducing fragmentation and UAF risks [8], [24]. Alternative strategies, including deferred memory reclamation methods (`call_rcu`) or rate-limiting VF churn, may also help but introduce complexity [1], [2].

Intel® has stated that this issue is not a security vulnerability because it requires root privileges to trigger. However, this work examines edge-case scenarios where kernel instability and memory exhaustion could still occur under normal operational conditions. As rapid provisioning and deprovisioning of Virtual Functions (VFs) could lead to severe memory fragmentation. Additionally, workloads involving frequent container or VM restarts, dynamic reconfiguration of SR-IOV devices, or network stress testing could escalate the risk of out-of-memory (OOM) conditions [13], [29], [30] Addressing this issue aligns with a defense-in-depth strategy to enhance kernel stability, even in privileged execution contexts [22].

Overall, our study highlights the importance of carefully managing memory reclamation in kernel operations [4]. Correctness and system stability should take precedence over minor performance gains achieved by asynchronous reclamation, particularly in administrative kernel paths like VF management [5], [16]. Findings advocate for robust synchronization mechanisms to prevent fragmentation, memory exhaustion, and use-after-free vulnerabilities, thereby improving overall system reliability.

## V. CONCLUSION & FUTURE WORK

Hash table management in the Linux kernel must carefully pair removals with appropriate RCU synchronization to avoid leaving behind ghost entries or overloading the memory subsystem [1], [2], [8], [12]. Through the case study of the Intel® ICE driver's VF handling, results show that the absence of `synchronize_rcu()` (or an equivalent mechanism) can cause two major issues: a fleeting period of stale pointers after deletion, and a tendency for unbounded memory allocation when operations are rapid, leading to OOM conditions [8], [12]. In our experiments, rapidly cycling VFs without RCU grace periods caused the kernel to temporarily retain dozens of VF structures and associated resources, eventually exhausting memory and triggering the OOM killer even though substantial free memory remained, which infers fragmentation-related exhaustion [2], [12], [16].

The remedy is clear: introducing a `synchronize_rcu()` call during VF teardown would ensure a clean quiescent state before freeing memory, thereby preventing stale lookups and pacing the teardown rate to what the system can handle [2]. This change, along with mindful memory management, restored stability in our tests as no OOM occurred when an artificial `synchronize_rcu()` was added in the loop, as expected.

Alternative mitigations, such as deferring frees with `call_rcu` or adding explicit rate limits, are secondary options but come with their trade-offs and complexity [1],[2]. The simplest and most robust solution is to wait for the RCU grace period on VF deletion [4]. This aligns with best practices followed in other kernel subsystems, where similar bugs were fixed by adding the missing synchronization barrier [3], [4]. Beyond the specific driver, our work serves as a reminder for kernel developers: when using RCU, always consider the lifecycle of your objects. Think about what happens if an object is created and destroyed in quick succession and test those scenarios.

The VF churn test is analogous to stress-testing other subsystems, such as rapidly adding and removing network interfaces or mounting and unmounting file systems in a loop, to ensure no lurking RCU issues. In conclusion, the lack of `synchronize_rcu()` in the ICE driver's VF hash table management causes severe memory fragmentation during rapid VF churn, often in the form of an OOM. By adding proper RCU synchronization or using deferred freeing correctly, the driver can prevent stale entries and keep memory usage in check [9], [12]. This yields a more reliable system that can handle even extreme cases gracefully [2]. All findings and recommendations were forwarded to the maintainers of the ICE driver. Going forward, we hope that this insight will help improve the driver and serve as a case study for the importance of RCU patterns in all forms of kernel development. Each subsystem should evaluate whether it has similar patterns and ensure that `synchronize_rcu()` (or analogous synchronization) is used whenever needed to balance RCU's deferred nature with timely cleanup, thereby maintaining consistency and preventing resource leaks in the face of concurrent operations [3].

## ACKNOWLEDGMENTS

Huge thanks to John Barry of Intel® for his supervision of the final year project this paper stemmed from. Thanks to Intel® PSIRT and Bug Bounty Program for timely responses and ease of communication and disclosure.

# This is a bit of an Anti-Project

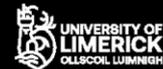

# General Introduction - What is my project about?

*Product Security? What is the Intel ICE Driver?*

- A **kernel-space** driver for the Intel **E810 Ethernet Controller(NIC)** hosted on Github.

- Manages network traffic, **virtual functions (VFs),** and **hardware resource allocation**.

- Why study its security?

- Vulnerabilities in NIC drivers can lead to system crashes, data breaches, and **denial-of-service (DoS) attacks**.

- Security flaws can affect **multi-tenant cloud environments** and **enterprise networks.**

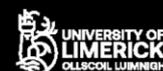

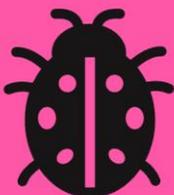

# Demo of Vulnerability

```
for ((;;));

do echo 0 > /sys/class/net/<Device-Name>/device/sriov_numvfs

&

echo 64 > /sys/class/net/<Device-Name>/device/sriov_numvfs

& done
```

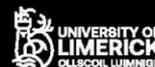

# 1. Demo: Denial Of Service – Exploiting RCU-Based VF Management Issues

- *Continuously creating and deleting VFs caused kernel memory exhaustion.*
- *Result: System crashed due to OOM killer activating.*

- *Presumed Root Cause: Delayed memory cleanup due to missing synchronize_rcu().*

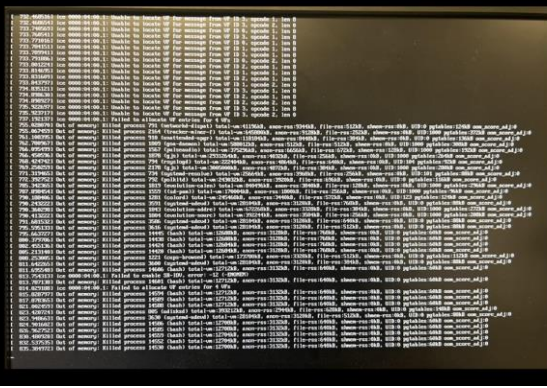

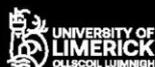

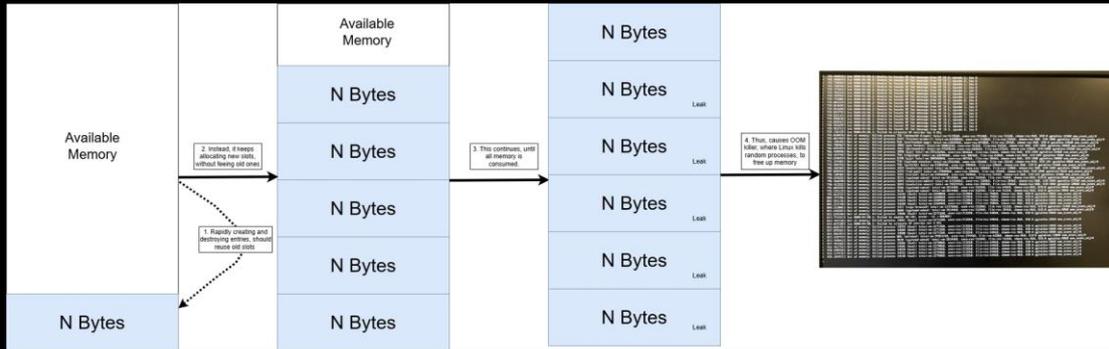
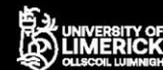
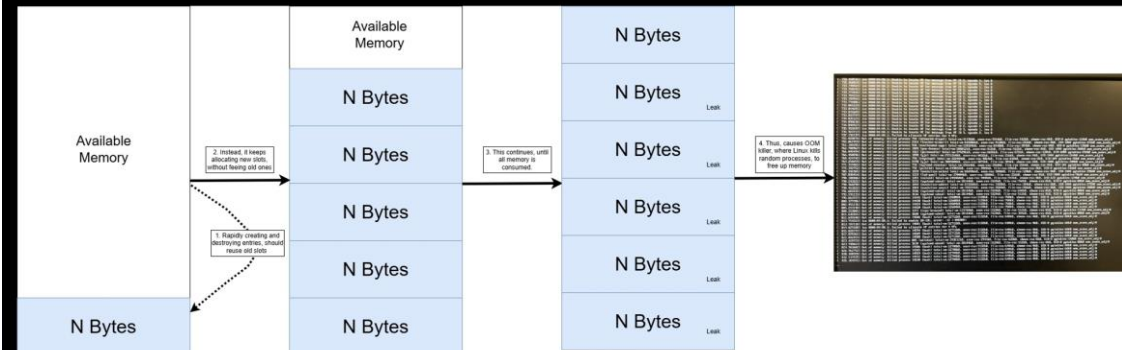
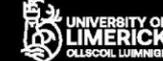

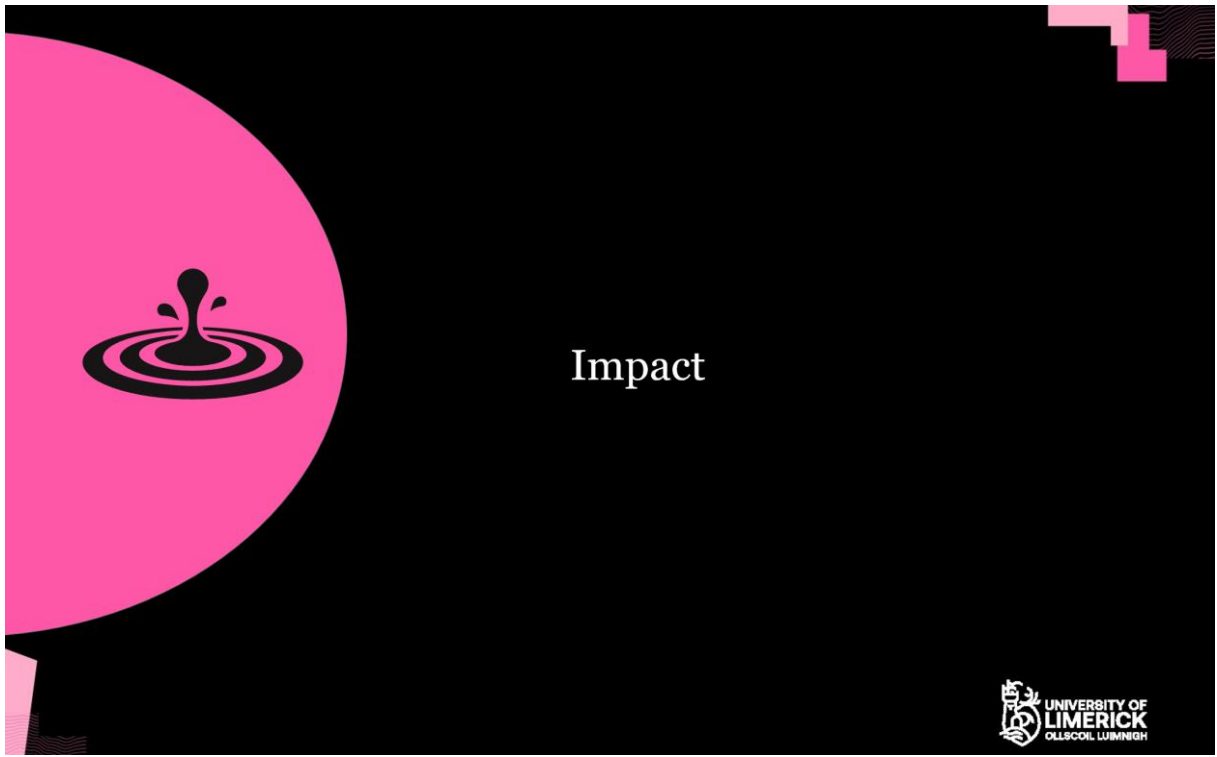

# 2. Impact

- *Intel's Ethernet product is used in millions of multi-tenant, data centres worldwide!*

- *The ICE driver is the primary driver for all of them! Incl. DPDK*

- *Thus, the impact that a privileged user can force a system reboot and/or instability, is huge.*

- *The presence of memory leaks exacerbates this further, as this <u>signed</u> driver could be used to break KASLR.*

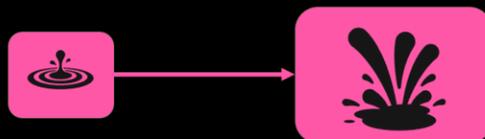

## 2. Impact

- Let's say, Intel has 51% of the NIC data centre market.

- 11,800 data centres worldwide
  [https://www.statista.com/chart/24149/data-centers-per-country/]

- ~200 million servers worldwide

- This affects an estimated 100 million servers worldwide!!!

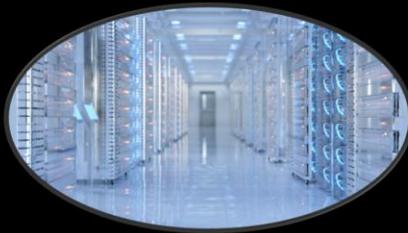

---

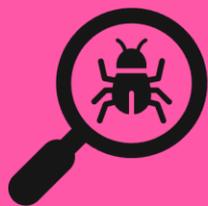

Discovery

# 3. Virtual Function Side-Channel Hash Table Tests – Introduction

**What is a side channel?**

*Often disputed, but to quote a contributor on **Spectre, Meltdown & KernelSnitch** "Using metadata to infer "secrets" about a target" i.e. Pizza Meter:*

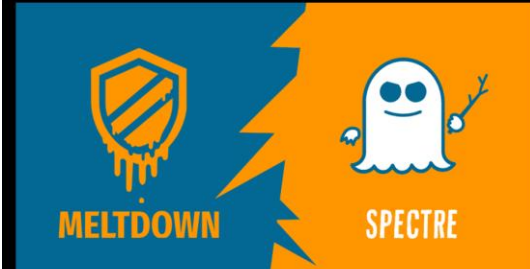
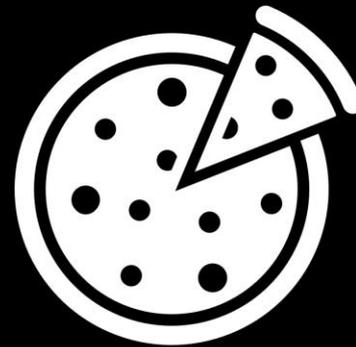
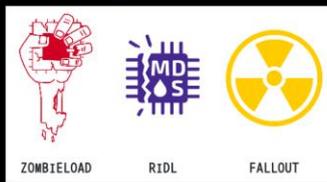
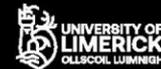

---

# 1. Discovery: Why was I timing in the first place?

*Side Channel Analysis:*

*KernelSnitch: A side-channel attack technique that exploits timing variations in kernel data structures [31]*

*Hash tables: A data structure used in the Linux kernel for fast lookups, storing elements in linked lists within indexed buckets*

*SR-IOV: A mechanism where a server, can allocate secure and isolated virtual NICs to it's VM tenants*

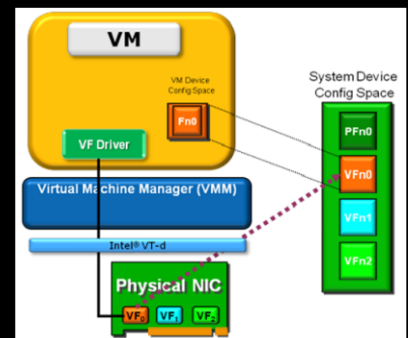
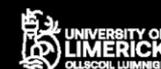

# 1. Discovery: What else did SCA show?

*Is it possible to infer hash table occupancy based on time?*

*Kernel Space Test:*                *User-Space Test:*

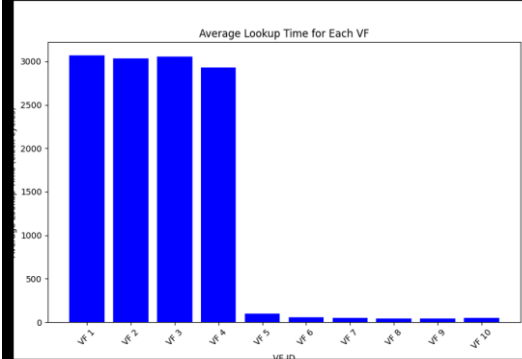 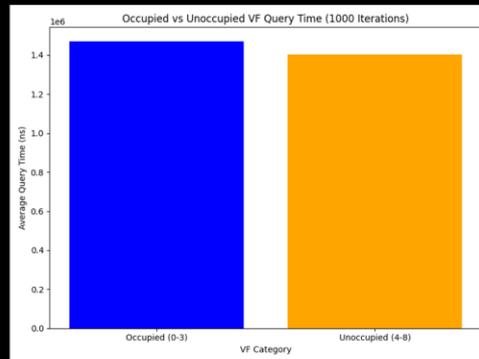

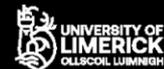

# 3. Discovery – Introduction

*This stems from what was thought to be a negative result:*

*I figured out that, after a VF is deleted, it maintains the timing value associated with an occupied slot for a short window.*

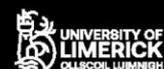

# 3. Discovery– Introduction

*Thus, I wanted to see how long a VF entry held onto the timing value associated with an occupied slot*

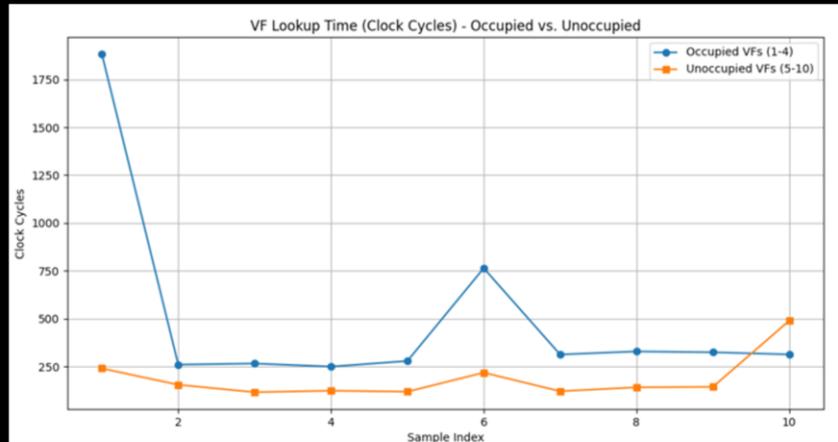

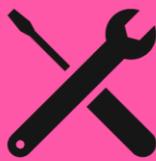

# Mitigations

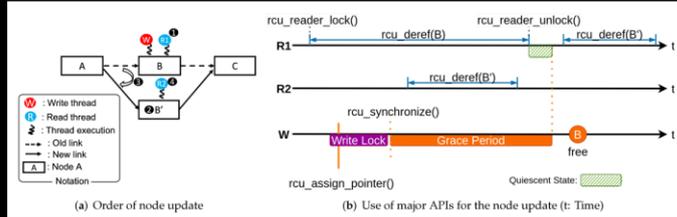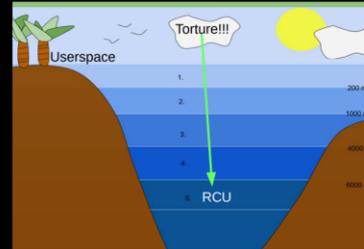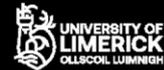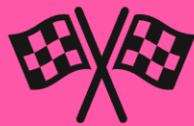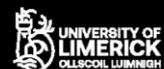

# Appendices

# 5. Appendices - Impact – Proof

- **Estimating Global Server Count:**

1. **Enterprise Data Centers** (~50% of total):
    1. Typically house **500 - 5,000 servers** each.
    2. Assuming an average of **2,000 servers per data center**.
    3. Estimated total: 11,800 × 50% × 2,000 = 11.8 million servers.
2. **Cloud & Hyperscale Data Centers** (~15% of total):
    1. These include AWS, Google Cloud, and Microsoft Azure, with **10,000 - 100,000 servers per facility**.
    2. Assuming an average of **50,000 servers per data center**.
    3. Estimated total: 11,800 × 15% × 50,000 = 88.5 million servers.
3. **Colocation Data Centers** (~30% of total):
    1. Typically house **100 - 1,000 servers** each.
    2. Assuming an average of **500 servers per data center**.
    3. Estimated total: 11,800 × 30% × 500 = 1.77 million servers.

4. **Mega Data Centers** (~5% of total):
    1. These large-scale facilities (like Facebook, Oracle, and Apple data centers) house **100,000 - 500,000 servers** each.
    2. Assuming an average of **200,000 servers per data center**.
    3. Estimated total: 11,800 × 5% × 200,000 = 118 million servers.

**Total Estimated Server Count:**
- Summing these up:
- **Enterprise Data Centers**: ~11.8 million
- **Cloud & Hyperscale Data Centers**: ~88.5 million
- **Colocation Data Centers**: ~1.77 million
- **Mega Data Centers**: ~118 million

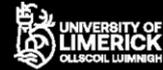

# Appendix C: Test Repository ReadMe

# FYP-Fuzzing-Suite

## Intel ICE Driver Security Audit Tools

This repository contains Python scripts developed for security analysis of Intel's ICE Driver for the E810 Ethernet Controller, focusing on fuzz testing and side-channel analysis.

## Files Overview

### #AdminQCmd-UI-Fuzz.py

**Purpose:** Performs fuzz testing on the Admin Queue (AdminQ) interface of the Intel ICE driver.
**Functionality:** Generates random and anomalous command inputs, evaluates driver robustness, integrates feedback-driven fuzzing, and monitors kernel logs for security vulnerabilities.

### #AdminQcmd-Fuzz-Test.py

**Purpose:** Specifically targets Admin Queue commands for fuzz testing.
**Functionality:** Executes randomized fuzzed commands against Admin Queue, analyzes kernel responses for errors, segmentation faults, and abnormal behaviors.

### #Access-Time-Average.py

**Purpose:** Analyzes and averages timing measurements during VF ID access tests.
**Functionality:** Computes average execution times for occupied vs. unoccupied VF lookups, highlighting timing-based side-channel vulnerabilities.

### #Debugfs_inputV_Fuzz_test.py

**Purpose:** Conducts fuzz testing on the driver's debugfs interface.
**Functionality:** Writes malicious payloads (large/malformed inputs) to debugfs files, monitors file integrity, kernel logs, and checks for potential crashes or improper handling.

### #adqsetup-atheris-instrumentation.py

**Purpose:** Implements fuzz testing for Adaptive Queueing (ADQ) configuration using Google's Atheris framework.
**Functionality:** Mutates ADQ configuration inputs, validates ICE driver's robustness against malicious or unexpected scenarios, ensures command stability.

### #ip-link-show-vf-id-test-regression.py

**Purpose:** Performs regression testing on VF IDs using Linux `ip link show`.
**Functionality:** Measures and compares query times for occupied vs. unoccupied VF IDs, assessing timing differences indicating VF occupancy status.

### #lspci-regression-test-vf.py

**Purpose:** Conducts timing-based side-channel testing using the `lspci` utility.
**Functionality:** Compares query times between valid and incorrect PCI addresses, identifies timing discrepancies exploitable to infer PCI device occupancy.

### #vf_occupancy_analyzer.py

**Purpose:** Analyzes kernel logs (`dmesg`) to classify VF occupancy status.
**Functionality:** Parses logs for VF lookup timings, categorizes VFs as occupied, unoccupied, or uncertain based on average clock cycles, exposing timing-based information leakage.

## CVSS Score

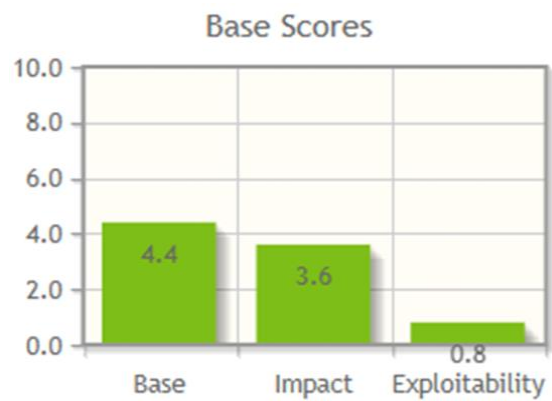

# Appendix E: Project Plan & Gantt chart

This action plan is structured around the key milestones highlighted in the Gantt chart, detailing each phase and corresponding tasks within the project's timeline. The project follows a phased approach to systematically evaluate the security of the Intel Columbiaville E810 Ethernet drivers, leveraging fuzzing techniques and vulnerability analysis.

**Milestone 1**: Preliminary Research and Setup (Completion: Early October)
This initial milestone focuses on establishing a strong foundation for the project by conducting background research, reviewing existing literature, and exploring fuzzing techniques. During this phase, a review of Intel's ICE driver GitHub repository will also be conducted.

**Key Tasks:**

- Background Research: Gain an understanding of Intel E810 driver functionality and existing security challenges.

- Literature Review: Compile and review relevant academic papers and industry reports on driver fuzzing, hardware security, and vulnerability mitigation techniques.

- Fuzzing Techniques Review: Explore and document various fuzzing methodologies (e.g., input-aware fuzzing, stateful fuzzing) and their applicability to the E810 driver.

- Exploration of Intel ICE Driver Repository: Examine Intel's publicly available driver code to understand potential vulnerabilities and gather insights into the structure of the E810 drivers.

*Progress: 100% complete as per Gantt chart*

**Milestone 2:** Interim Report Formulation and Experiments (Completion: Mid-November)
This milestone is crucial as it involves setting up the hardware and dependencies needed for experimentation and formulating a detailed plan for fuzzing and vulnerability testing. This is also when the interim report on initial findings and progress is prepared.

**Key Tasks:**

- Set Up Hardware and Dependencies: Install and configure the necessary hardware (servers, network interfaces) and software dependencies required for testing the drivers. A couple of Intel e810s and cables have been supplied by intel Shannon and are set up in 2 PCs in b2005.

- Driver Build and Compilation: Build the Intel E810 drivers from source to create an isolated test environment for fuzzing and security analysis.

- Plan Formulation: Finalize the experimental plan, defining the testing methodology, expected outcomes, and evaluation criteria for vulnerabilities.

- Report Writing: This includes gathering references, formulating an abstract, writing the literature review, and finalizing the experiment plan for the interim report.

**Milestone 3**: Code Review & Fuzzing Campaign (Late October to February)

The core of the project takes place in this phase, where a comprehensive fuzzing campaign is conducted. The goal is to uncover vulnerabilities through systematic testing of the Intel E810 drivers.

**Key Tasks:**

- Create Fuzzing Harnesses: Build custom fuzzing harnesses that allow the fuzzer to interact effectively with the E810 drivers.

- Corpus Setup for Fuzzing: Create input corpuses, including valid and edge-case scenarios, to ensure comprehensive fuzzing coverage.

- Fuzzing Initiation: Begin fuzzing tests on the E810 drivers, continuously monitoring and adjusting test parameters to improve vulnerability detection rates.

- Hardware Review and Monitoring: Ensure that hardware settings are optimized for fuzzing and that the system's performance is adequately monitored to avoid crashes.

- Documentation Review: Conduct an in-depth review of Intel's documentation on driver security features, ensuring compliance and identifying potential gaps.

**Milestone 4:** Documentation Review and Vulnerability Analysis (January to March)

This phase involves analysing the vulnerabilities uncovered during fuzzing and reviewing the security implications of any findings. It includes both technical analysis and reporting of the weaknesses found.

**Key Tasks:**

- Crash and Hang Analysis: Examine any crashes or hangs encountered during fuzzing to determine if they reveal security vulnerabilities.

- Triage of Vulnerabilities: Assess the severity of discovered vulnerabilities, determining which ones can be exploited and require immediate action.

- Previous Mitigation Analysis: Review and analyse any previously implemented security measures (e.g., Secure Boot) to evaluate their effectiveness.

- Documentation of Findings: Maintain comprehensive documentation of all vulnerabilities and exploits found, categorized by risk level.

- Implication Review: Analyse the broader implications of each vulnerability in terms of network security, data integrity, and system performance.

*Progress: Scheduled for early 2024 according to Gantt chart timeline*

**Milestone 5:** Final Reporting and Presentation (April 2024)

The final phase involves compiling all findings, developing mitigation strategies, and presenting the results in a comprehensive report. This report will detail all vulnerabilities discovered, their potential impact, and recommended solutions.

**Key Tasks:**

- Scope Legality: Review the legal implications of fuzzing and vulnerability disclosure to Intel, ensuring compliance with ethical guidelines.

- Final Report Formulation: Document the overall findings, vulnerability analysis, and mitigation recommendations in a detailed report.

- Presentation Preparation: Prepare a formal presentation summarizing the project outcomes, highlighting key vulnerabilities, and offering actionable recommendations for improving driver security.

*Progress: Set to begin in late March 2024*

By following this extended action plan, the project will ensure that the Intel Columbiaville E810 Ethernet drivers undergo thorough security testing, with findings documented and addressed in a systematic manner. The project will result in actionable insights into potential vulnerabilities, providing Intel with the necessary recommendations for securing their driver software. Each phase is aligned with the project timeline, ensuring that progress is closely monitored and adjusted as needed.

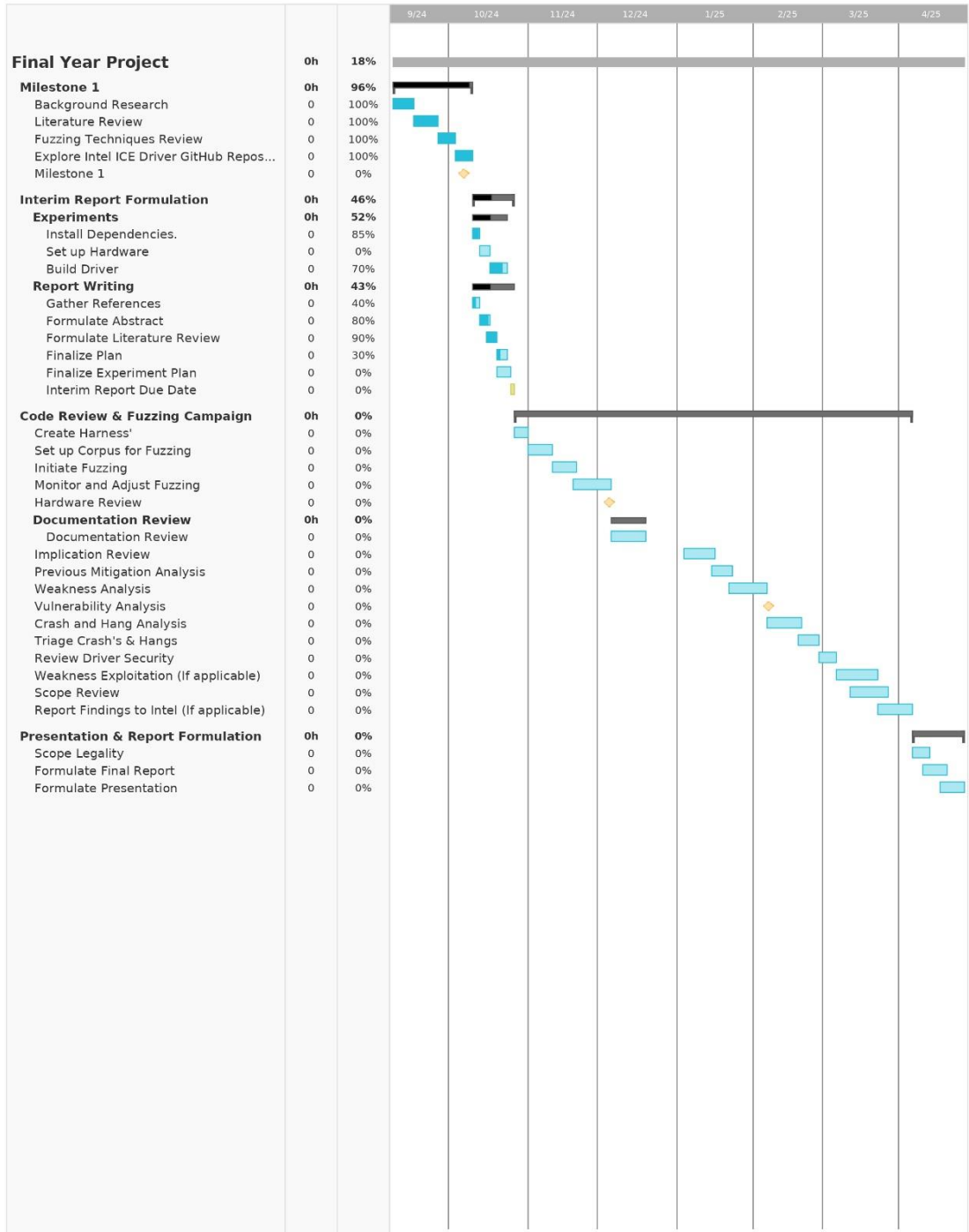